\documentclass{article}

\usepackage{xspace}
\usepackage{authblk}
\usepackage{amsthm} % for proof envirnment

\usepackage{siunitx}
\usepackage{url}
\usepackage{hyperref}
\hypersetup{
  colorlinks=true,
  linkcolor=blue,
  filecolor=magenta,
  urlcolor=cyan,
  pdftitle={Overleaf Example},
  pdfpagemode=FullScreen,
}

\usepackage{coollist}
\usepackage{xparse}

\ExplSyntaxOn
\DeclareExpandableDocumentCommand{\eval}{m}{\int_eval:n {#1}}
\ExplSyntaxOff

\usepackage{fancyvrb}
\usepackage{enumitem}
\usepackage{listings}
\usepackage{pgffor}
\usepackage{caption}
\usepackage[table]{xcolor}
\usepackage{graphicx}
\usepackage{ifthen}
\usepackage[top=1in,bottom=1in]{geometry}  % See geometry.pdf to learn the
\usepackage{adjustbox}
\usepackage{array}

\newcolumntype{R}[2]{%
    >{\adjustbox{angle=#1,lap=\width-(#2)}\bgroup}%
    l%
    <{\egroup}%
}
\usepackage{setspace}
\usepackage{relsize}
\newenvironment{narrow}[2]{%
  \begin{list}{}{%
      \setlength{\topsep}{0pt}%
      \setlength{\leftmargin}{#1}%
      \setlength{\rightmargin}{#2}%
      \setlength{\listparindent}{\parindent}%
      \setlength{\itemindent}{\parindent}%
      \setlength{\parsep}{\parskip}}%
  \item[]}{\end{list}}

\newcommand\FigDiff[1]{Figure~\ref{#1} on page~\pageref{#1}}
\newcommand\FigSame[1]{Figure~\ref{#1}}
\newcommand\Figref[1]{\ifthenelse{\value{page}=\pageref{#1}}
  {\FigSame{#1}}{\FigDiff{#1}}}

\usepackage{float}

\usepackage{amsmath}

\usepackage{tikz-cd}
\usepackage{listings}

\newcommand*{\R}{\ensuremath{\mathbb{R}}\xspace}

\begin{document}

\def\spacingset#1{\renewcommand{\baselinestretch}%
{#1}\small\normalsize} \spacingset{1}

\title{A New Approach to Compositional Data Analysis using \(L^{\infty}\)-normalization with Applications to Vaginal Microbiome}
\author[1]{Pawel Gajer\thanks{Research reported in this publication was supported by grant number INV-048956 from the Gates Foundation.}}
\author[1]{Jacques Ravel}
\affil[1]{Center for Advanced Microbiome Research and Innovation (CAMRI), Institute for Genome Sciences, and Department of Microbiology and Immunology, University of Maryland School of Medicine}
\date{\vspace{0.1cm}\today}
\date{\today}

\maketitle

\bigskip
\begin{abstract}
  This paper introduces a novel approach to compositional data analysis based on
  \(L^{\infty}\)-normalization, addressing the challenges posed by zero-rich
  high-throughput compositional data such as microbiome datasets. Traditional
  methods like Aitchison's logistic transformations require excluding zeros,
  which conflicts with the reality that most high-throughput omics datasets
  contain structural zeros that cannot be removed without violating the inherent
  structure of the system. Moreover, such datasets exist exclusively on the
  boundary of compositional space, making traditional interior-focused
  approaches fundamentally misaligned with the data's true nature.

  We present a family of \(L^{p}\)-normalizations, focusing primarily on
  \(L^{\infty}\)-normalization due to its advantageous properties. This approach
  identifies the compositional space with the \(L^{\infty}\)-simplex, which can
  be represented as a union of top-dimensional faces called
  \(L^{\infty}\)-cells. Each \(L^{\infty}\)-cell consists of samples where one
  component's absolute abundance equals or exceeds all others, and carries a
  coordinate system identifying it with a d-dimensional unit cube.

  When applied to vaginal microbiome data, the \(L^{\infty}\)-decomposition
  method aligns well with established Community State Type (CST) classifications
  while offering several advantages: (1) each \(L^{\infty}\)-CST is named after
  its dominating component, (2) \(L^{\infty}\)-decomposition has clear
  biological meaning, (3) it remains stable under addition or subtraction of
  samples, (4) it resolves issues associated with cluster-based approaches, and
  (5) each \(L^{\infty}\)-cell provides a homogeneous coordinate system for
  exploring internal structure.

  Furthermore, we extend homogeneous coordinates to the entire sample space
  through a cube embedding technique, mapping compositional data into a
  d-dimensional unit cube. These various cube embeddings can be integrated
  through their Cartesian product, providing a unified representation of
  compositional data from multiple perspectives. While demonstrated through
  microbiome studies, these methods are applicable to any compositional data
  type.
\end{abstract}

\noindent%
\vfill

\section*{1. Introduction}

The advent of high-throughput techniques, generating mainly compositional data,
has introduced unique analytical challenges. Aitchison's groundwork in
compositional data analysis, which proposed logistic transformations
necessitating exclusion of zeros
\cite{aitchison1982statistical,aitchison1986book}, contrasts starkly with the
zero-rich nature of high-throughput compositional data. From a geometric
viewpoint, assuming data has no zeros implies that it resides entirely within
the interior of the compositional space. However, in reality, most high-throughput
omics datasets exist solely on the boundary of the compositional space, with not a
single data point located in the space's interior. To address this issue,
researchers have turned to methods such as missing-data imputation or the
addition of pseudo-counts to eliminate zeros, effectively shifting the data from
the compositional space's boundary to its interior. These adjustments have a
considerable impact on the outcomes of subsequent analyses, especially
considering the prevalence of zeros in most high-throughput datasets. Despite
the significance of these changes, conducting sensitivity analyses to evaluate
their effects is remarkably neglected.

To address the challenges posed by zeros in compositional data, we introduce a
family of \(L^{p}\)-normalizations, with \(p\) representing either a positive
real number or infinity. The Total Sum Scaling (division of the rows of a data
matrix by the total row sums), used routinely to normalize compositional data,
is a special case of the family of transformations corresponding to \(p = 1\).
Our focus, however, is primarily on \(L^{\infty}\)-normalization due to its
advantageous properties.

\(L^{\infty}\)-normalization identifies the compositional space with the
\(L^{\infty}\)-simplex, which can be represented as a union of its
top-dimensional faces, called \(L^{\infty}\)-cells (see Figure~\ref{fig1:fig}).
The \(k\)-th cell consists of samples where the absolute abundance of the
\(k\)-th component is equal to or greater than the absolute abundances of all
other components. For example, in the context of shotgun metagenomic data, the
\(k\)-th component consists of samples where the abundance of the \(k\)-th gene
is greater than or equal to the abundance of any other genes. Moreover, each
cell carries a coordinate system identifying it with the \(d\)-dimensional unit
cube \([0,1]^{d}\), where \(d\) is the dimensionality of the compositional space
equal to the number of components minus one. Although high-throughput datasets
might feature a large number of components, corresponding to different types of
biomarkers, and consequently, \(L^{\infty}\)-cells, the count of
\(L^{\infty}\)-cells that actually hold data tends to be notably small.

\begin{center}
  \begin{narrow}{-0.3in}{0in}
    \includegraphics[scale=0.6]{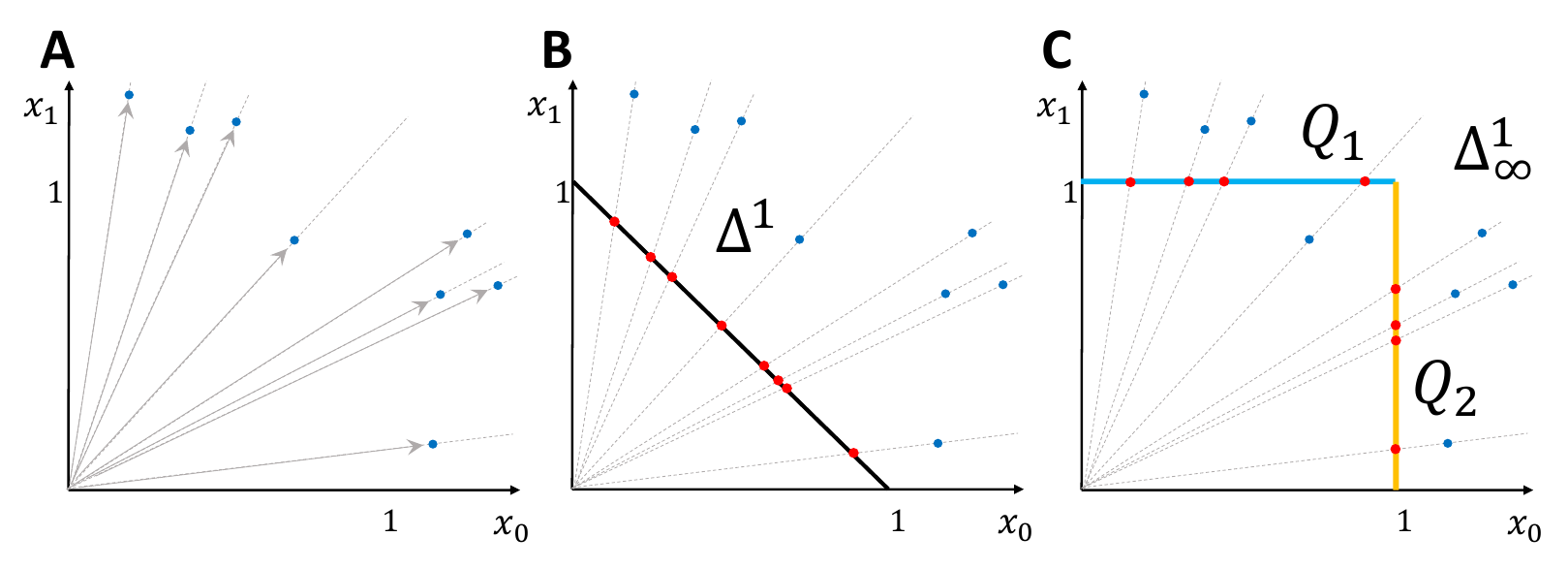}
  \end{narrow}
  \captionof{figure}{Compositional data reflects relative proportions, not
    absolute values. A compositional measurement vector (an arrow with a blue
    disk at the end in panel A) represents any point along the dashed line, as
    they share the same proportions. This leads to multiple ways to represent
    the compositional space as a hypersurface. Panel B shows the standard
    simplex (\(\Delta^{1}\)) representation of 1D compositional space, while
    panel C shows the \(L^{\infty}\)-simplex (\(\Delta^{1}_{\infty}\))
    representation, together with its two top dimensional faces
    (\(L^{\infty}\)-cells): \(Q_{1}\) in blue and \(Q_{2}\) in orange. The
    faces \(Q_{1}\) and \(Q_{2}\) can be identified with the unit interval
    \([0,1]\). In higher dimensions, each \(L^{\infty}\)-cell can be identified
    with a cube \([0,1]^{d}\), where \(d\) is the dimensionality of the
    compositional space equal to the number of components minus
    one. \label{fig1:fig}}
\end{center}
In studies involving human vaginal micobiome, analyzing unique DNA sequence
variations — known as Amplicon Sequence Variants (ASVs) — using the
\(L^{\infty}\)-decomposition method aligns well with established methods for
categorizing vaginal microbial communities into different community types. This
suggests that \(L^{\infty}\)-decomposition could be a viable alternative method
for high-level characterization of these and other bacterial communities as well
as other compositional data types.

Since, from the perspective of downstream analyses it is practical to focus on
\(L^{\infty}\)-cells with a substantial number of data samples, we define
truncated \(L^{\infty}\)-decomposition for the compositional data, reassigning
samples from less populated \(L^{\infty}\)-cells to those with adequate number
of samples. This rearrangement is supported by the observation that samples in
lesser-populated \(L^{\infty}\)-cells are typically situated near the boundary
of that cell adjoining \(L^{\infty}\)-cells with a high sample count. In this
context, we refer to the cells of truncated \(L^{\infty}\)-decomposition as
\(L^{\infty}\)-CSTs, providing a new perspective on community state
classifications.

\(L^{\infty}\)-CSTs have several advantages over the classical CSTs or
enterotypes (in the context of gut microbiome): 1) The name of each
\(L^{\infty}\)-CST is the name of the component that dominates (at the absolute
abundance) the corresponding set of samples, 2) the definition of
\(L^{\infty}\)-CST has a simple and easy to understand biological meaning, 3)
\(L^{\infty}\)-decomposition is stable under addition or subtraction of samples
- that is the membership of a sample in a given \(L^{\infty}\)-cell is not
dependent on other samples, 4) \(L^{\infty}\)-cells are not clusters, but
absolute abundances dominance patters of the components of the data - resolving
all issues associated with the construction of CSTs or enterotypes as clusters
(see Section~6 for more details), 5) \(L^{\infty}\)-cell is not only a grouping
of samples, but each \(L^{\infty}\)-cell comes with a homogeneous coordinate
system that allows further elucidation of the internal structure of that cell.
This projective geometry coordinate system, introduced by August Ferdinand
Möbius in 1827, corresponds to Cartesian coordinates in Euclidean geometry
\cite{mobius1827barycentrische}. The log-transform of homogeneous coordinates is
the Aitchison's additive log ratio transform.

Utilizing elementary geometro-topological ideas, we have extended homogeneous
coordinates from a subset of samples, where the denominator of the
transformation is non-zero, to the entire sample space. This expansion results
in a new parametrization of the compositional data, which we refer to as a cube
embedding, that maps the data into a \(d\)-dimensional cube \([0,1]^{d}\), where
\(d\) is the number of components of the compositional data minus one.

Cube embeddings result in a variety of compositional data representations,
providing insights into the data's structure from multiple perspectives. Similar
to how varying angles of tomographic imaging are employed to piece together the
three-dimensional structure of internal organs, or how different maps of the
Earth facilitate analysis of different geo-spacial phenomena, this technique
facilitates a more comprehensive understanding of compositional data's
structure. However, the abundance of representations introduces complexity,
prompting the question: Can these diverse representations be integrated? A
unified representation of the data can be taken to be the Cartesian product of
the cube embeddings associated with \(L^{\infty}\)-CSTs. This approach is
detailed in Section~10.

The emphasis in the paper is on compositional data in the context of microbiome
studies. Yet, all methods can be applied in the context of any type of
compositional data.

The paper is organized as follows: Section 2 introduces the concept of
compositional spaces and their properties. Section 3 discusses subcompositional
coherence and its relevance to omics data. Section 4 explores various
parametrizations of compositional spaces, focusing on \(L^p\)-normalizations.
Section 5 presents the \(L^{\infty}\)-decomposition of compositional spaces into
\(L^{\infty}\)-cells. Section 6 compares VALENCIA Community State Types (CSTs)
with \(L^{\infty}\)-CSTs, illustrating the advantages of the latter. Section 7
describes the alignment of \(L^{\infty}\)-cells through rotation to create a
global coordinate system. Section 8 introduces hypercube embeddings of
compositional data, extending homogeneous coordinates to the entire sample
space. Section 9 demonstrates the integration of cube embeddings to provide a
unified representation of the data. Section 10 concludes with a discussion of
the implications and potential applications of the presented methods.

%%% Local Variables:
%%% mode: latex
%%% TeX-master: "Linf_paper"
%%% End:
 % 1
\section*{2. Compositional Spaces}

Most omics data types, such as 16S rRNA and metagenomic, exhibit an
asymptotically compositional nature. This means that for two read count sets
\(x = (x_{0}, x_{1}, \ldots, x_{d})\) and
\(x' = (x'_{0}, x'_{1}, \ldots, x'_{d})\) from the same sample, with total
read counts \(T(x)\) and \(T(x')\) respectively, as the total read counts
\(T(x)\) and \(T(x')\) increase, the proportions \(\frac{x}{T(x)}\) and
\(\frac{x'}{T(x')}\) become closer and closer to each other. That is
\[
  \lim_{\min(T(x), T(x')) \rightarrow \infty} \bigg\| \frac{x}{T(x)} - \frac{x'}{T(x')} \bigg\| = 0.
\]
The deviations from exact proportionality can be attributed to sampling errors
and, more significantly, the presence of zero components due to detection
limits.

In practical applications, it is assumed that the data is compositional, meaning
that a vector representing proportions of different components in a sample is
defined up to a positive scaling factor. Since all components, \(x_{i}\), of a
compositional vector \(x = (x_{0}, x_{1}, \ldots, x_{d})\) are non-negative
and the vector cannot be zero, it can be identified with a point of
\(\R^{d+1}_{\ge 0} - \{0\}\). In this context, any compositional vector \(x\)
represents a class of equivalent vectors:
\[
  [x] = \{\lambda x: x \in \R^{d+1}_{\ge 0} - \{0\}, \lambda > 0\}
\]
derived from scaling \(x\) by various factors. The equivalence class of a point
\((x_{0}, x_{1}, \ldots , x_{d})\) is denoted as
\([x_{0}: x_{1}: \ldots : x_{d}]\). Geometrically, \([x]\) represents the line
passing through the origin in \(\R^{d+1}_{\ge 0}\) spanned by \(x\) as
illustrated in Figure~\ref{comp.2d.ex1:fig}.
\begin{center}
\includegraphics[scale=0.2]{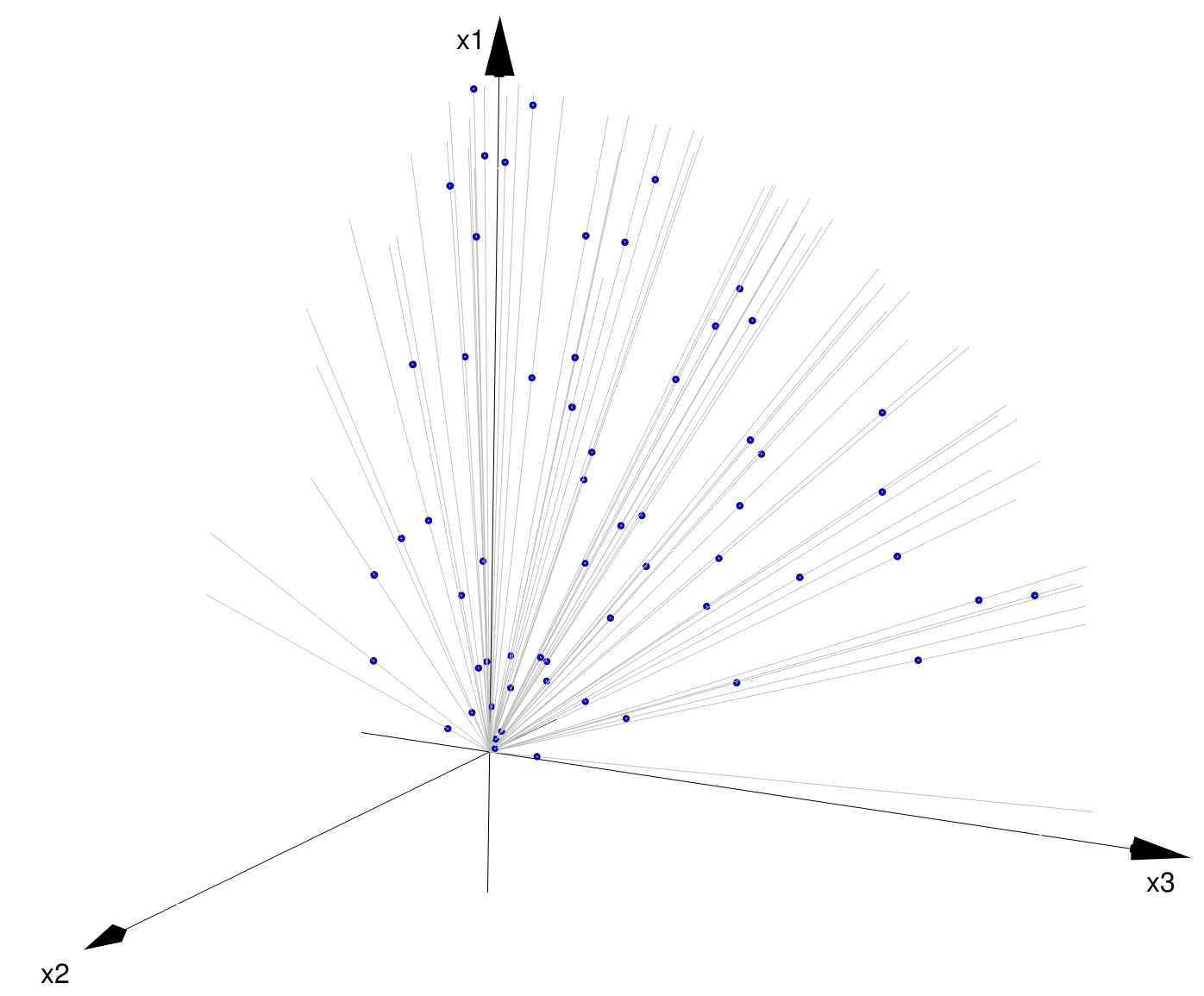}
\captionof{figure}{Random sample of points of \(\mathbb{RP}^{2}_{\ge 0}\)
  represented by lines through the origin with representatives that determine
  these lines marked with blue spheres.
  \label{comp.2d.ex1:fig}}
\end{center}

The space \(\mathbb{RP}^{d}_{\ge 0}\) of lines through the origin in
\(\R^{d+1}_{\ge 0}\), hereafter referred to as the \emph{\(d\)-dimensional
  compositional space}, is a subset of a \emph{real projective space}
\(\mathbb{RP}^{d}\) defined as a space of lines through the origin in
\(\R^{d+1}\). Since every line through the origin in \(\R^{d+1}\) intersects the
unit sphere \(S^{d}\) in exactly two antipodal points, \(\mathbb{RP}^{d}\) can
be identified with the quotient space of the sphere \(S^{d}\) with every pair of
antipodal points collapsed to a single point. Real projective space is a
fundamental example of a non-trivial smooth manifolds
\cite{o2006elementary,guillemin2010differential}. That is, there does not exist
a global coordinate system on \(\mathbb{RP}^{d}\) that would identify that space
with \(\mathbb{R}^{d}\). Instead, \(\mathbb{RP}^{d}\) is equipped with an atlas
of charts, \(\{\phi_{\alpha}: U_{\alpha} \rightarrow \R^{d}\}_{\alpha \in I}\),
such that every point of \(\mathbb{RP}^{d}\) belongs to at least one
\(U_{\alpha}\) and if \(U_{\alpha} \cap U_{\beta} \ne \emptyset\), then the
composition \(\phi_{\alpha} \circ \phi_{\beta}^{-1}: \R^{d} \rightarrow \R^{d}\)
is a diffeomorphism, which means that it is a smooth map whose inverse is also
smooth \cite{guillemin2010differential}. The standard atlas of
\(\mathbb{RP}^{d}\) consists of homogeneous coordinate charts. The \(i\)-th
homogeneous coordinate chart is a map
\(\phi_{i}: U_{i} = \mathbb{RP}^{d} - \{x_{i} = 0\} \rightarrow \R^{d}\) defined
as
\[
  \phi_{i}([x_{0}: x_{1}: \ldots : x_{d}]) = (\frac{x_{0}}{x_{i}}, \ldots, \frac{x_{i-1}}{x_{i}}, 1, \frac{x_{i+1}}{x_{i}}, \ldots, \frac{x_{d}}{x_{i}})
\]
where \(\{x_{i} = 0\}\) is a subset of \(\mathbb{RP}^{d}\) consisting of points
\([x_{0}: x_{1}: \ldots : x_{d}]\) of such that \(x_{i} = 0\). Thus, each
point in \(\{x_{i} = 0\}\) corresponds to a line in \(\R^{d+1}\) contained in
the hyperplane \(x_{i} = 0\). Geometrically, the map \(\phi_{i}\) assigns to the
line \([x]\), span by a vector \(x \in \R^{d+1} - \{0\}\), the point of
intersection of \([x]\) with the hyperplane \(x_{i} = 1\). Going forward, the
notation \(\phi_{i}\) will also refer to the restriction of this map to the
compositional space \(\mathbb{RP}^{d}_{\ge 0} \subset \mathbb{RP}^{d}\).
\begin{center}
\includegraphics[scale=0.6]{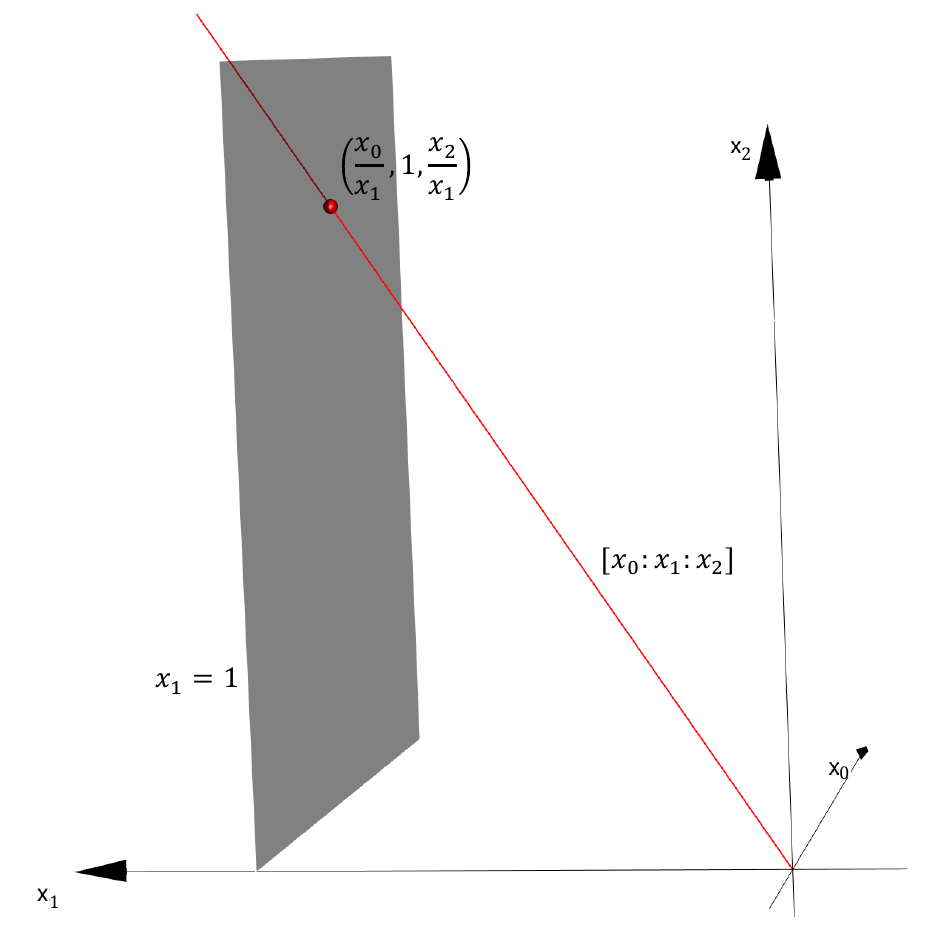}
\captionof{figure}{Geometric interpretation of the \(x_{1}\)-homogeneous
  coordinate chart \(\phi_{1}\) as a map that assigns to the line
  \(\{\lambda (x_{0}, x_{1}, x_{2})\}\), shown in red, representing a point
  \([x_{0}: x_{1}: x_{2}]\) of \(\mathbb{RP}^{d}\), the intersection
  \(\big(\frac{x_{0}}{x_{1}}, 1, \frac{x_{2}}{x_{1}}\big)\) of the line with the
  plane \(x_{1} = 1\). }
\end{center}

A function \(f: \mathbb{RP}^{d}_{\ge 0} \rightarrow \R\) over a compositional
space \(\mathbb{RP}^{d}_{\ge 0}\) assigns a unique real number \(f([x])\) to
each point \([x]\) within \(\mathbb{RP}^{d}_{\ge 0}\). Since,
\(\mathbb{RP}^{d}_{\ge 0}\) is a quotient of \(\R^{d+1}_{\ge 0} - \{0\}\), and
we typically use representatives \((x_{0}, x_{1}, \ldots, x_{d})\) of points
in \(\mathbb{RP}^{d}_{\ge 0}\) in practical applications, it’s important to
characterize a function over \(\mathbb{RP}^{d}_{\ge 0}\) in terms of
\(\R^{d+1}_{\ge 0} - \{0\}\). A function
\(f: \R^{d+1}_{\ge 0} - \{0\} \rightarrow \R\) induces a function over
\(\mathbb{RP}^{d}_{\ge 0}\) if it is scale invariant, which means that
\(f(x) = f(\lambda x)\) holds true for all \(x\) in \(\R^{d+1}_{\ge 0} - \{0\}\)
and any positive value of \(\lambda\).

\textit{Example 2-A:} Let
\(d: \mathbb{RP}^{d}_{\ge 0} \times \mathbb{RP}^{d}_{\ge 0} \rightarrow [0, \infty)\)
be a metric over \(\mathbb{RP}^{d}_{\ge 0}\), then for any point
\(x_{\text{ref}}\) of \(\mathbb{RP}^{d}_{\ge 0}\), the distance to
\(x_{\text{ref}}\), \(f(x) = d(x, x_{\text{ref}})\), is a function over
\(\mathbb{RP}^{d}_{\ge 0}\).

\textit{Example 2-B:} The angular distance
\[
  d_{\text{angle}}([x_{0}: \ldots : x_{d}], [y_{0}: \ldots : y_{d}]) = \cos^{-1}\big( \left\langle \pi_{1}([x_{0}: \ldots : x_{d}]), \pi_{1}([y_{0}: \ldots : y_{d}]) \right\rangle \big)
\]
where
\[
  \pi_{1}([x_{0}: \ldots : x_{d}]) = \frac{(x_{0},\ldots, x_{d})}{\| (x_{0},\ldots, x_{d}) \|_{1}}
\]
assigns to each line
\(\{\lambda (x_{0}, \ldots, x_{d})\}_{\lambda \in [0,\infty)}\) representing
\([x_{0}: \ldots : x_{d}]\), the unit vector on that line. The inner product
between two unit vectors is the cosine of the angle between them. Thus, the
angular distance between two points of \(\mathbb{RP}^{d}_{\ge 0}\) is the angle
between the associated unit vectors.

The angular distance is an example of a construct of a metric on
\(\mathbb{RP}^{d}_{\ge 0}\) that is induced by a metric on a particular
parametrization of \(\mathbb{RP}^{d}_{\ge 0}\). In the case of angular distance
it is an \(L^{2}\) or spherical parametrization of \(\mathbb{RP}^{d}_{\ge 0}\).
Different parametrizations of compositional spaces will be discussed in more
detail in Section~3. Thus, even if a dissimilarity measure is not well defined
on a compositional spaces, any parametrization of a compositional space can be used
to extend the dissimilarity measure from the parametrization to the compositional
space. For example, Bray-Curtis dissimilarity \(d_{\text{BC}}\) is not well
defined on \(\mathbb{RP}^{d}_{\ge 0} \times \mathbb{RP}^{d}_{\ge 0}\), as it is
not homogeneous, meaning
\(d_{\text{BC}}(\lambda_{0}x, \lambda_{1}y) = d_{\text{BC}}(x, y)\) is not true
for \(d_{\text{BC}}\) for any \(\lambda_{0}, \lambda_{1} > 0\). Yet, the
restriction of \(d_{\text{BC}}\) to a particular parametrization \(\pi\) of
\(\mathbb{RP}^{d}_{\ge 0}\) defines a dissimilarity measure on
\(\mathbb{RP}^{d}_{\ge 0}\):
\[
  d_{\text{BC}}([x], [y]) :=  d_{\text{BC}}(\pi([x]), \pi([y])).
\]

\textit{Example 2-C:} The ratio of any two components
\([x_{0}: x_{1}: \ldots : x_{d}] \mapsto \frac{x_{j}}{x_{i}}\) is a function
\(\phi_{ji}: \mathbb{RP}^{d}_{\ge 0} - \{x_{i} = 0\} \rightarrow [0, \infty)\)
over \(\mathbb{RP}^{d}_{\ge 0} - \{x_{i} = 0\}\), where \(\{x_{i} = 0\}\) is a
subset of points \([x_{0}: x_{1}: \ldots : x_{d}]\) of
\(\mathbb{RP}^{d}_{\ge 0}\) with \(x_{i} = 0\), as for any representative
\((\lambda x_{0}, \lambda x_{1}, \ldots , \lambda x_{d})\) of
\([x_{0}: x_{1}: \ldots : x_{d}]\), we have
\(\frac{\lambda x_{j}}{\lambda x_{i}} = \frac{x_{j}}{x_{i}}\).

\textit{Example 2-D:} In the one-dimensional case, the function
\(\phi_{1}: \mathbb{RP}^{1}_{\ge 0} - \{x_{1} = 0\} \rightarrow [0, \infty)\) is
defined by \(\phi_{1}([x_{0}: x_{1}]) = \frac{x_{0}}{x_{1}}\). \(\phi_{1}\)
extends to a function
\(\hat{\phi}_{1}: \mathbb{RP}^{1}_{\ge 0} \rightarrow [0, \infty)*\), where
\(\hat{\phi}_{1}([1:0]) = \infty\), and \([0, \infty)*\) is a one-point
compactification of \([0, \infty)\), that adds a point \(\infty\) to the
original interval with the open intervals \((a, \infty)\) forming open
neighborhoods of \(\infty\) in \([0, \infty)*\)
\cite{mendelson1990introduction}. This compactified space \([0, \infty)*\) is
homeomorphic to the interval \([0,1]\). That is, it is continuous and has an
inverse that is also continuous \cite{mendelson1990introduction}. In fact, any
continuous and monotonically increasing function
\(\sigma: [0, \infty) \rightarrow [0,1)\) that approaches 1 as \(x\) approaches
infinity, induces a homeomorphism between \([0, \infty)*\) and \([0,1]\) if
extended by setting \(\sigma(\infty) = 1\). For example, we can take \(\sigma\)
to be \(\sigma(x) = 1 - e^{- \lambda x}\) for any \(\lambda > 0\).

Similarly, a function
\(\phi_{1}^{\sigma} = \sigma \circ \phi_{1}: \mathbb{RP}^{1}_{\ge 0} - \{x_{1} = 0\} \rightarrow [0, 1)\)
defined as
\(\phi_{1}^{\sigma}([x_{0}: x_{1}]) = \sigma\left(\frac{x_{0}}{x_{1}}\right)\),
extends to a function
\(\hat{\phi}_{1}^{\sigma}: \mathbb{RP}^{1}_{\ge 0} \rightarrow [0, 1]\), where
\(\hat{\phi}_{1}^{\sigma}([1: 0]) = 1\). This extension is a homeomorphism
between \(\mathbb{RP}^{1}_{\ge 0}\) and the unit interval \([0,1]\). Later in
this section, we will explore a higher-dimensional generalization of this
function that allows for the parametrization of \(\mathbb{RP}^{d}_{\ge 0}\) via
the unit hyper-cube \([0,1]^{d}\).
\begin{center}
  \begin{narrow}{-0.3in}{0in}
    \includegraphics[scale=0.65]{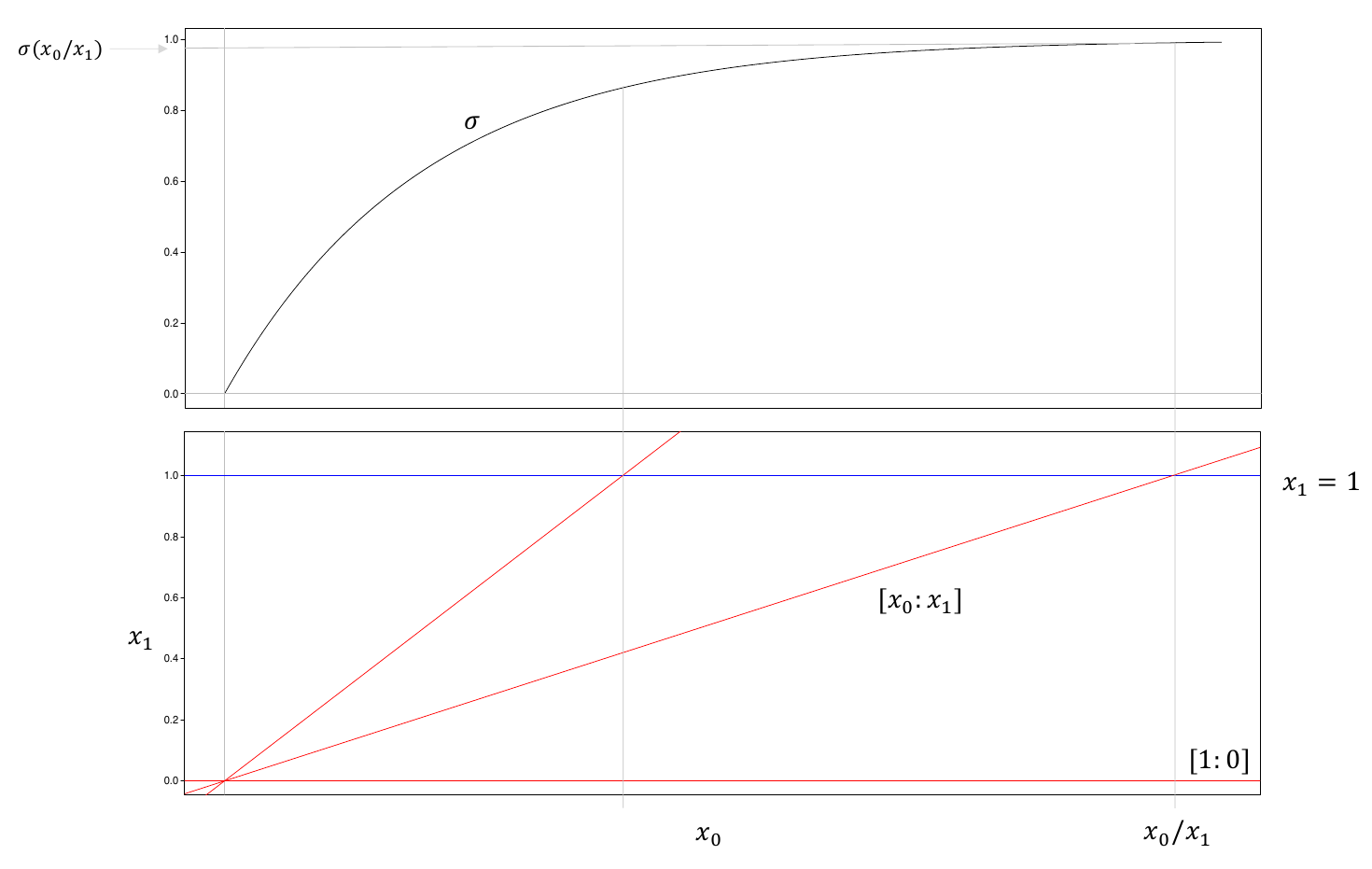}
  \end{narrow}
  \captionof{figure}{Extension of \(\phi_{1}^{\sigma}\) to
    \(\mathbb{RP}^{1}_{\ge 0}\). Each red line in the plot of the lower panel
    corresponds to a point of \(\mathbb{RP}^{1}_{\ge 0} - \{ x_{1} = 0\}\). The
    \(x\)-coordinate of the intersection of each of the red lines with the blue
    line of \(x_{1} = 1\), is the geometric interpretation of the map
    \(\phi_{1}\). The vertical gray lines indicate the values of \(\phi_{1}\) at
    the corresponding points of \(\mathbb{RP}^{1}_{\ge 0} - \{ x_{1} = 0\}\). As
    the point (red line) \([x_{0}:x_{1}]\) of
    \(\mathbb{RP}^{1}_{\ge 0} - \{ x_{1} = 0\}\) gets closer and closer to the
    line \(x_{1} = 0\), corresponding to the point \([1:0]\) of
    \(\mathbb{RP}^{1}_{\ge 0}\), the value of \(\sigma\) at
    \(\phi_{1}([x_{0}:x_{1}])\) gets closer and closer to 1. Thus, settting
    \(\phi_{1}^{\sigma}([1:0]) = 1\) extends \(\phi_{1}^{\sigma}\) to a mapping
    identifying \(\mathbb{RP}^{1}_{\ge 0}\) with \([0,1]\).  }
\end{center}

\textit{Example 2-E:} The \(i\)-th coordinate \(x_{i}\) of a point
\([x_{0}: x_{1}: \ldots : x_{d}]\) is not a function over the compositional space
\(\mathbb{RP}^{d}_{\ge 0}\) as for two different representations
\((x_{0}, x_{1}, \ldots , x_{d})\) and
\((\lambda x_{0}, \lambda x_{1}, \ldots , \lambda x_{d})\) of
\([x_{0}: x_{1}: \ldots : x_{d}]\), \(i\)-th coordinates \(x_{i}\) and
\(\lambda x_{i}\) of these representations have distinct values for
\(\lambda \ne 1\). Thus, if we take \(\mathbb{RP}^{d}_{\ge 0}\) to be a sample
space, the \(i\)-th coordinate assignment
\([x_{0}: x_{1}: \ldots : x_{d}] \mapsto x_{i}\) is not random variable over
\(\mathbb{RP}^{d}_{\ge 0}\).

The notion of a function over \(\mathbb{RP}^{d}_{\ge 0}\) extends to the notion
of a mapping \(f: \mathbb{RP}^{d}_{\ge 0} \rightarrow X\) from
\(\mathbb{RP}^{d}_{\ge 0}\) to any topological space \(X\). Such mapping assigns
exactly one value to every point \([x]\) of \(\mathbb{RP}^{d}_{\ge 0}\).
Similarly, as in the function case, a mapping
\(f: \R^{d+1}_{\ge 0} - \{0\} \rightarrow X\) induces a mapping
\(f': \mathbb{RP}^{d}_{\ge 0} \rightarrow X\) if \(f\) is scale invariant. A
mapping \(f: \mathbb{RP}^{d}_{\ge 0} \rightarrow X\) is a parametrization of
\(\mathbb{RP}^{d}_{\ge 0}\) if it is a homeomorphism. We will explore different
parametrizations of \(\mathbb{RP}^{d}_{\ge 0}\) in Section~4.

%%% Local Variables:
%%% mode: latex
%%% TeX-master: "Linf_paper"
%%% End:
 % 2
\section*{3. Subcompositional Coherence}

Compositional data in omics studies, such as those found in 16S rRNA amplicon,
metabolomics, and metagenomics, inherently include only a subset of all possible
components. This subcompositional nature arises because some microbes may be
present at levels below detectable thresholds and thus remain undetected and
omitted from the dataset, despite their actual presence. This selective
exclusion of components based on detectability skews the relative abundances of
those detected. It is crucial for data analysis methods to acknowledge and
directly address this subcompositional character of the data.

In the context of omics data, component sub-sampling is not random but
inherently selective, focusing predominantly on the most reliably detectable
components. This aspect challenges the traditional definition of
subcompositional coherence, which demands that analytical methods yield
consistent results, irrespective of whether they are applied to the entire set
of components or just a subset. Traditionally, subcompositional coherence is
defined as the invariance of a procedure or transformation with respect to an
arbitrary operation of subsetting components. More formally, a family of maps
\(\{f^{d}: U^{d} \rightarrow \R^{n(d)}\}_{d > 0}\), where \(n(d)\) is a positive
integer function of \(d\) and \(U^{d}\) is a subset of
\(\mathbb{RP}^{d}_{\ge 0}\), is considered subcompositionally coherent if for
any component sub-setting operation \(\pi\), there exists a corresponding map
\(\pi'\) that maintains the coherence as illustrated by the following
commutative diagram:
\begin{center}
\begin{tikzcd}[column sep=huge, row sep=huge] U^d \arrow[r, "f^d"] \arrow[d,
"\pi"', left] & \mathbb{R}^{n(d)} \arrow[d, "\pi'"]\\ U^m \arrow[r, "f^m"',
below] & \mathbb{R}^{n(m)} \end{tikzcd}
\end{center}
However, this standard definition of subcompositional coherence requires the
procedure or transformation to be invariant with respect to any sub-sampling of
components operation, which may be overly stringent for omics data. In the
context of omics studies, a more appropriate criterion for subcompositional
coherence would be to maintain consistency when sub-sampling eliminates
components of low abundance. Therefore, for compositional omics data, a
transformation or procedure might only need to maintain consistency when less
abundant components are eliminated. This modified version of subcompositional
coherence, tailored to the selective nature of omics data sub-sampling, ensures
that the analytical methods yield consistent results when applied to either the
full compositional data or a subset of its components, focusing on the most
reliably detectable components.

\textit{Example 3-A:} The first component ratio charts
\(\{\phi_{0}^{d}: \mathbb{RP}^{d}_{\ge 0} - \{x_{0} = 0\} \rightarrow \mathbb{R}^{d}\}\)
\[
  \phi^{d}_{0}([x_{0}: x_{1}: x_{2}: \ldots : x_{d}]) = \bigg(\frac{x_{1}}{x_{0}}, \frac{x_{2}}{x_{0}}, \ldots, \frac{x_{d}}{x_{0}}\bigg)
\]
are subcompositionally coherent with respect to an arbitrary subsetting
operation as restriction of \(\phi^{d}_{0}\) to any subset of \((m+1)\)
components (that does not include \(x_{0}\)) gives the map \(\phi^{m}_{0}\). For
example, for \(d = 5\)

\[
  \phi^{4}_{0}([x_{0}: x_{1}: x_{2}: x_{3}: x_{4}]) = \bigg(\frac{x_{1}}{x_{0}}, \frac{x_{2}}{x_{0}}, \frac{x_{3}}{x_{0}}, \frac{x_{4}}{x_{0}}\bigg)
\]
is invariant with respect to the subset selection of the first three components
as then \(\phi^{5}_{0}\) becomes \(\phi^{3}_{0}\)
\[
  \phi_{0}^{2}([x_{0}: x_{1}: x_{2}]) = \bigg(\frac{x_{1}}{x_{0}}, \frac{x_{2}}{x_{0}}\bigg)
\]
Thus, \(\pi' \circ \phi_{0}^{4} = \phi_{0}^{2} \circ \pi\) for \(\pi\) the
sub-setting of the first three components and
\(\pi': \mathbb{R}^4 \rightarrow \mathbb{R}^2\) the projection on the first two
components \(\pi'(r_{0}, r_{1}, r_{2}, r_{3}) = (r_{0}, r_{1})\).

\textit{Example 3-B:} The centered ratio transformations
\(\{\psi^{d}: \mathbb{RP}^{d}_{\ge 0} - \{\prod_{i = 0}^{d}x_{i} = 0\} \rightarrow \mathbb{R}^{d}\}\)
\[
  \psi^{d}([x_{0}: x_{1}: \ldots : x_{d}]) = \bigg(\frac{x_{0}}{\big( \prod_{i = 0}^{d}x_{i} \big)^{1/(d+1)}},  \frac{x_{1}}{\big( \prod_{i = 0}^{d}x_{i} \big)^{1/(d+1)}}, \ldots, \frac{x_{d}}{\big( \prod_{i = 0}^{d}x_{i} \big)^{1/(d+1)}}\bigg)
\]
are not subcompositionally coherent. For example, for \(d = 5\)
\[
  \psi^{4}([x_{0}: x_{1}: x_{2}: x_{3}: x_{4}]) = \bigg(\frac{x_{0}}{\big( \prod_{i = 0}^{4}x_{i} \big)^{1/5}},  \frac{x_{1}}{\big( \prod_{i = 0}^{4}x_{i} \big)^{1/5}}, \frac{x_{2}}{\big( \prod_{i = 0}^{4}x_{i} \big)^{1/5}}, \frac{x_{3}}{\big( \prod_{i = 0}^{4}x_{i} \big)^{1/5}}, \frac{x_{4}}{\big( \prod_{i = 0}^{4}x_{i} \big)^{1/5}}\bigg)
\]
and if we restrict \(\psi^{4}\) to the first three components, we will not get
\(\psi^{2}\)
\[
  \psi^{2}([x_{0}: x_{1}: x_{2}]) = \bigg(\frac{x_{0}}{\big( \prod_{i = 0}^{2}x_{i} \big)^{1/3}},  \frac{x_{1}}{\big( \prod_{i = 0}^{2}x_{i} \big)^{1/3}}, \frac{x_{2}}{\big( \prod_{i = 0}^{2}x_{i} \big)^{1/3}}\bigg)
\]
as in general \(\frac{x_{i}}{\big( x_{0}x_{1}x_{2} \big)^{1/3}}\) is not equal
to \(\frac{x_{i}}{\big( x_{0}x_{1}x_{2}x_{3}x_{4} \big)^{1/5}}\). Thus, the
centered log ratio is not invariant with respect to the component sub-setting
operation. This implies that the the centered log ratio transformations are also
not subcompositionally coherent.

%%% Local Variables:
%%% mode: latex
%%% TeX-master: "Linf_paper"
%%% End:
 % 3
\section*{4. Parametrizations of Compositional Spaces} \label{param:sec}

There are infinite number of parametrizations of \(\mathbb{RP}^{d}_{\ge 0}\).
Indeed, any \(d\)-dimensional smooth hypersurface \(H\) in \(\R^{d+1}\) with the
property that each line, representing a point of \(\mathbb{RP}^{d}_{\ge 0}\),
intersects \(H\) at exactly one point, induces a parametrization of
\(\mathbb{RP}^{d}_{\ge 0}\). For example, if we take \(H\) to be the unit
\(d\)-dimensional sphere \(S^{d}\), we get a homeomorphism between
\(\mathbb{RP}^{d}_{\ge 0}\) and the \(L^{2}\)-simplex
\[
\Delta^{d}_{2} = S^{d} \cap \R^{d+1}_{\ge 0} = \{x \in \R^{d+1}_{\ge 0}: \| x\|_{2} =  1\}
\]
where \(\| x\|_{2} = \big( \sum_{i=0}^{d} |x_{i}|^{2} \big)^{1/2} \) is the
\(L^{2}\)-norm of \(x\). The projection on the unit sphere
\(\pi_{1}: \R^{d+1}_{\ge 0} - \{0\} \rightarrow \Delta^{d}_{2}\) defined by
\(\pi_{1}(x) = \frac{x}{\| x\|_{2}}\) is called the \(L^{2}\)-normalization of
the compositional data.

Similarly, any \(L^{p}\)-norm, \(p \ge 1\) or \(p = \infty\), or
\(L^{p}\)-quasi-norm, where \(0 < p < 1\), induces a parametrization of
\(\mathbb{RP}^{d}_{\ge 0}\) with the corresponding \(L^{p}\)-normalization
given by the projection
\(\pi_{p}: \R^{d+1}_{\ge 0} - \{0\} \rightarrow \Delta^{d}_{p}\), where
\(\Delta^{d}_{p}\) is the \(L^{p}\)-simplex
\[
\Delta^{d}_{p}  = \{x \in \R^{d+1}_{\ge 0}: \| x\|_{p} =  1\}
\]
where \(\pi_{p}(x) = \frac{x}{\| x\|_{p}}\) with
\(\| x\|_{p} = \big( \sum_{i=0}^{d} |x_{i}|^{p} \big)^{1/p} \) for \(p > 0\)
and \(\| x\|_{\infty} = \max_{i} |x_{i}|\). In particular, for \(p = 1\), the
standard simplex \(\Delta^{d}\) is the intersection of the \(L^{1}\) unit sphere
\[
  S^{d}_{1} = \{x \in \R^{d+1}:  \| x\|_{1} =  1\}
\]
and the positive orthant \(\R^{d+1}_{\ge 0}\).

The following figures show 1d and 2d unit \(L^{p}\)-spheres with the corresponding
\(L^{p}\) simplices (in red).
% \textbf{[Figures]}
\begin{center}
  %\begin{narrow}{-0.6in}{0in}
    \begin{minipage}{0.33\linewidth}
      \includegraphics[scale=0.31]{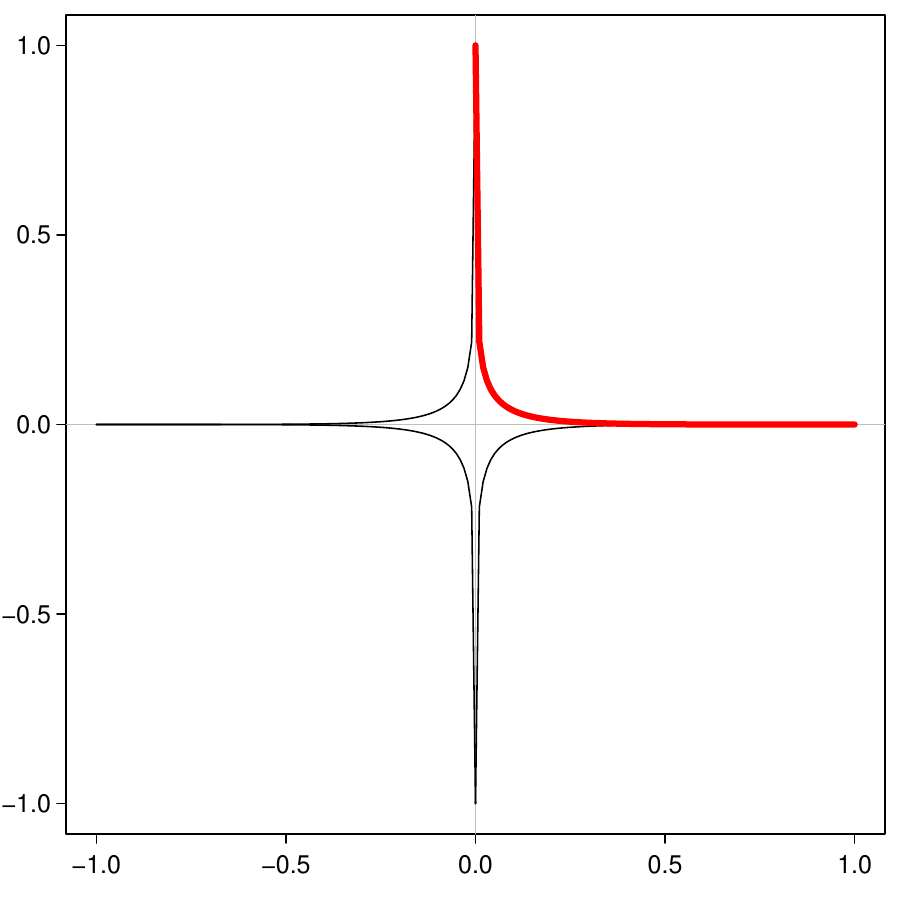}
    \end{minipage}%
    \hspace{0.15em}% Negative space here
    \begin{minipage}{0.33\linewidth}
      \includegraphics[scale=0.31]{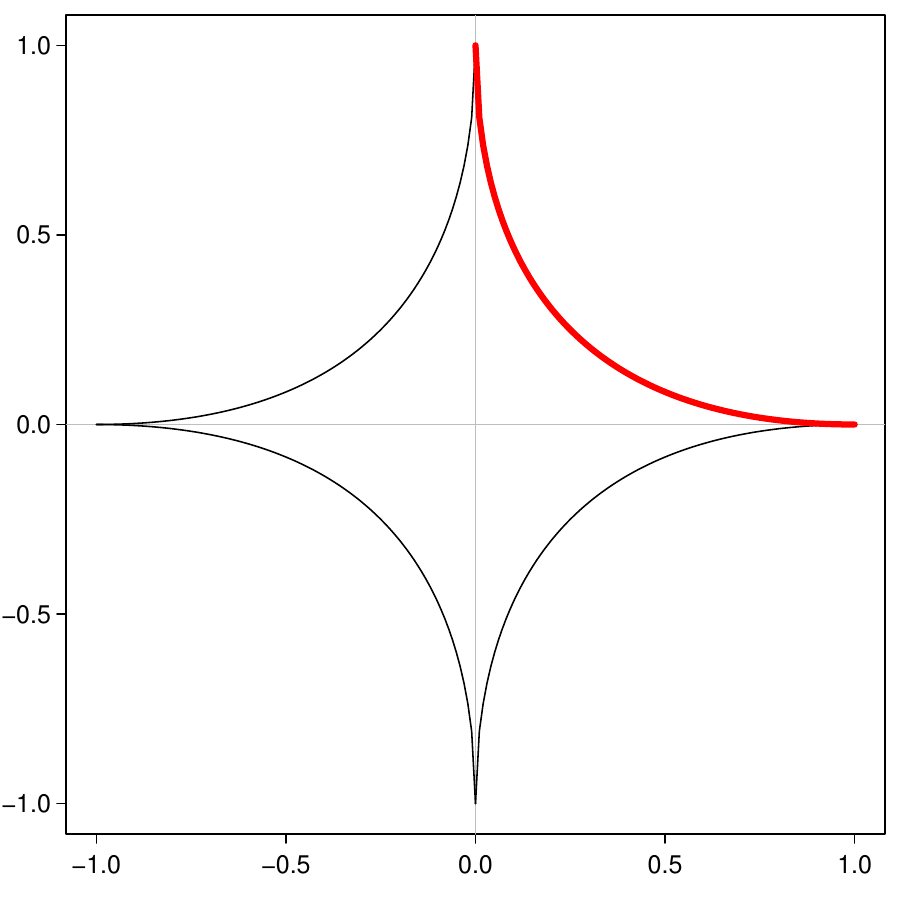}
    \end{minipage}%
    \hspace{0.15em}% Negative space here
    \begin{minipage}{.33\linewidth}
      \includegraphics[scale=0.31]{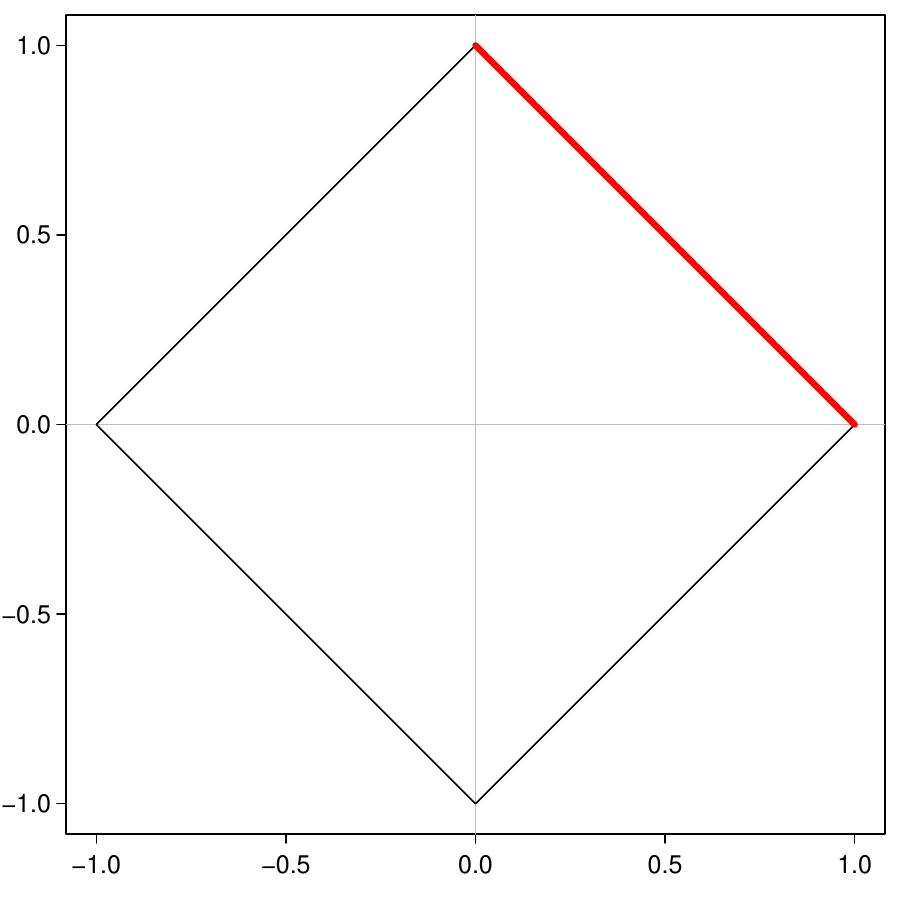}
    \end{minipage}
  %\end{narrow}
    \captionof{figure}{One dimensional unit \(L^{p}\)-spheres with the
      corresponding \(L^{p}\) simplices for \(p = 0.25\) (left) \(p = 0.5\)
      (middle) and \(p = 1\) (right). \label{Lp.1d.spheres.p1:fig}}
\end{center}

\begin{center}
  %\begin{narrow}{-0.6in}{0in}
    \begin{minipage}{0.33\linewidth}
      \includegraphics[scale=0.31]{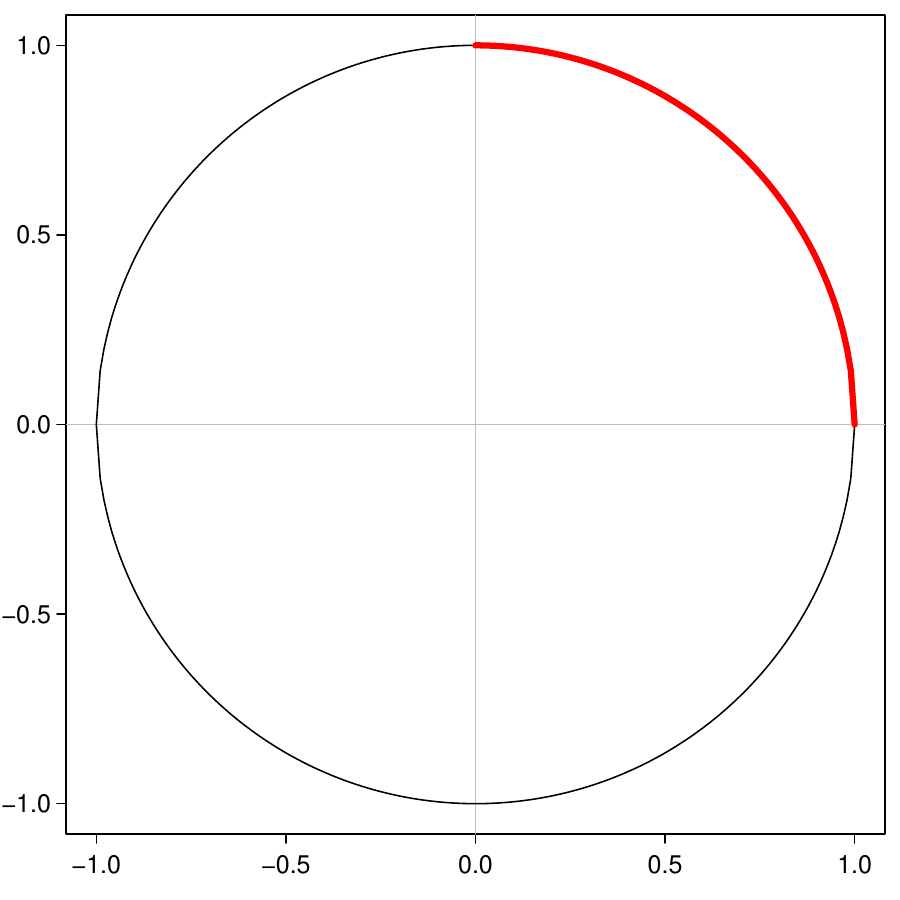}
    \end{minipage}%
    \hspace{0.15em}% Negative space here
    \begin{minipage}{0.33\linewidth}
      \includegraphics[scale=0.31]{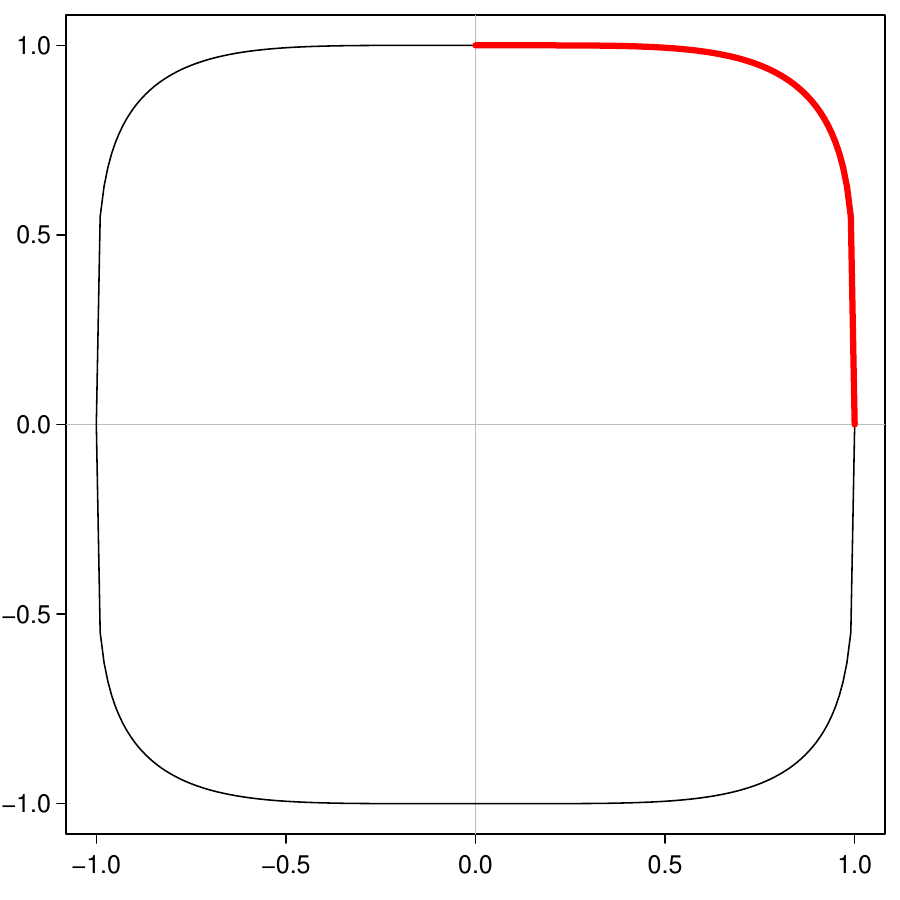}
    \end{minipage}%
    \hspace{0.15em}% Negative space here
    \begin{minipage}{.33\linewidth}
      \includegraphics[scale=0.31]{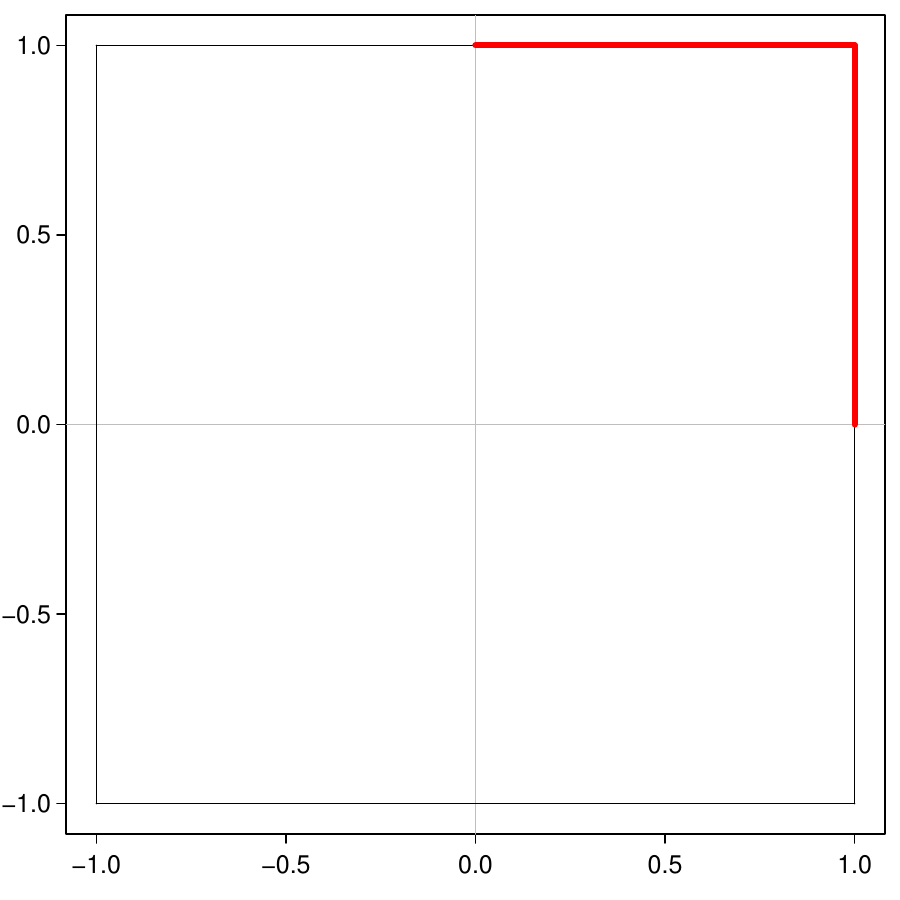}
    \end{minipage}
  %\end{narrow}
    \captionof{figure}{One dimensional unit \(L^{p}\)-spheres with the
      corresponding \(L^{p}\) simplices for \(p = 2\) (left) \(p = 5\) (middle)
      and \(p = \infty\) (right). \label{Lp.1d.spheres.p2:fig}}
\end{center}

\begin{center}
  %\begin{narrow}{-0.6in}{0in}
    \begin{minipage}{0.33\linewidth}
      \includegraphics[scale=0.11]{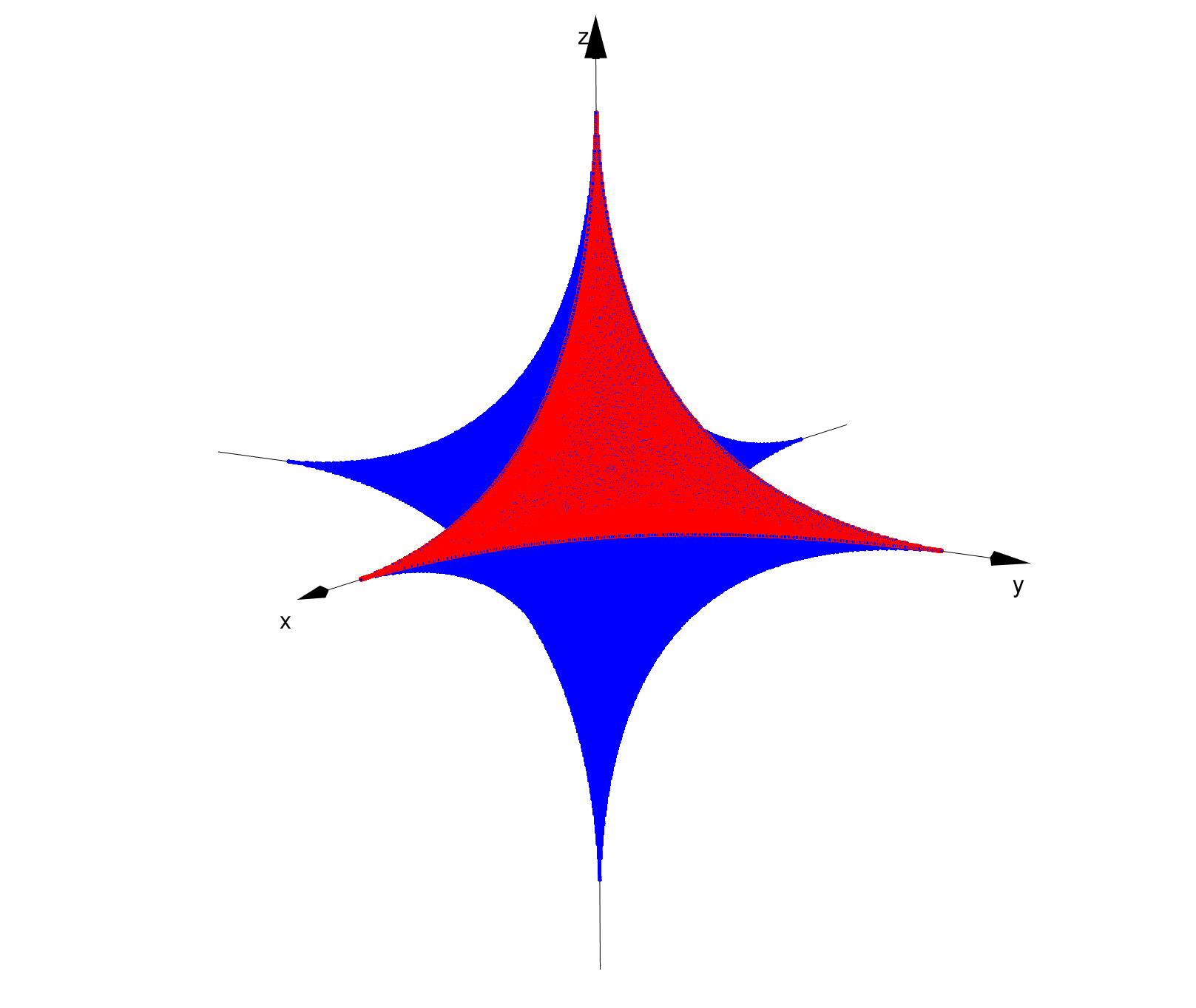}
    \end{minipage}%
    \hspace{0em}% Negative space here
    \begin{minipage}{0.33\linewidth}
      \includegraphics[scale=0.11]{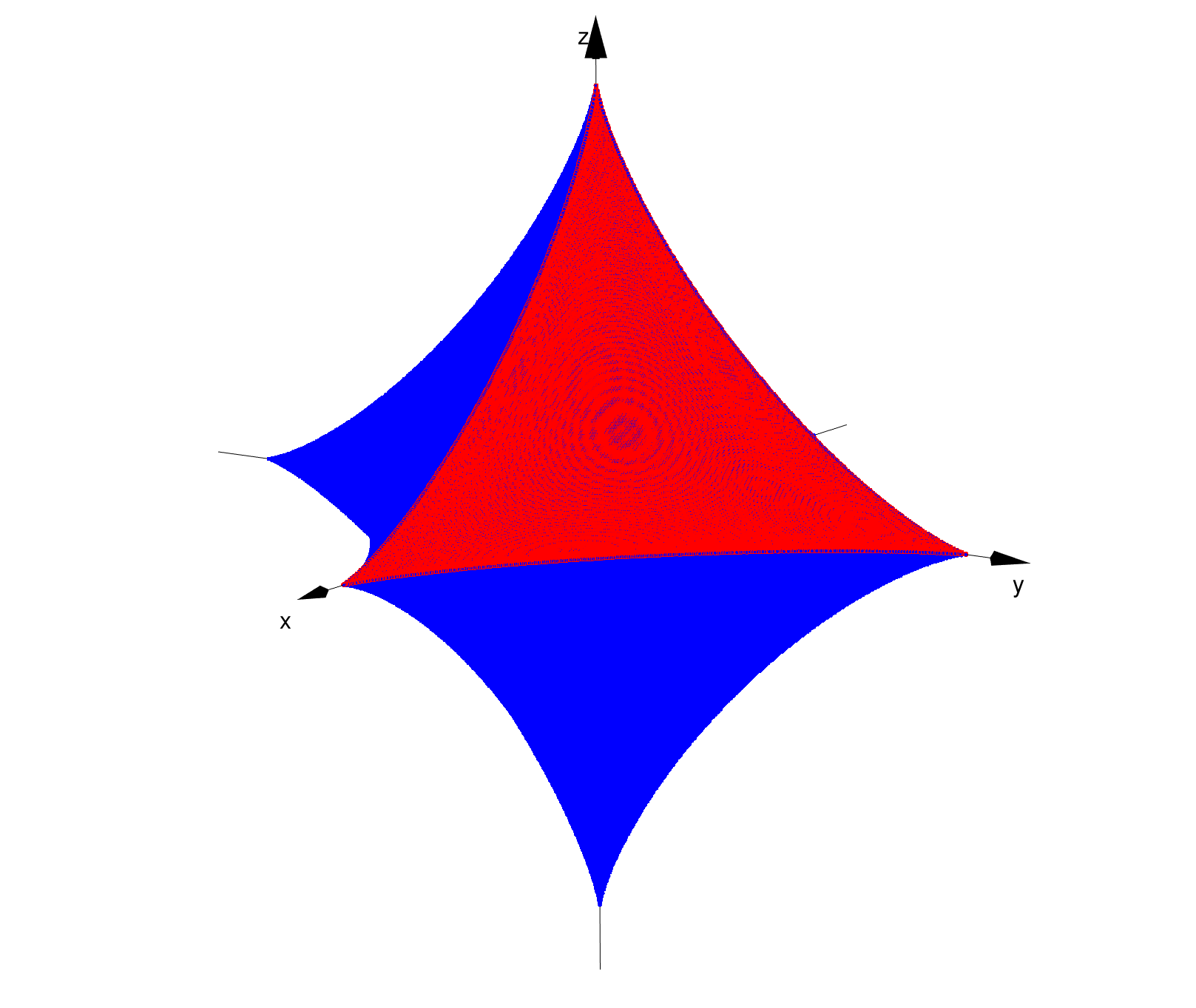}
    \end{minipage}%
    \hspace{0em}% Negative space here
    \begin{minipage}{.33\linewidth}
      \includegraphics[scale=0.11]{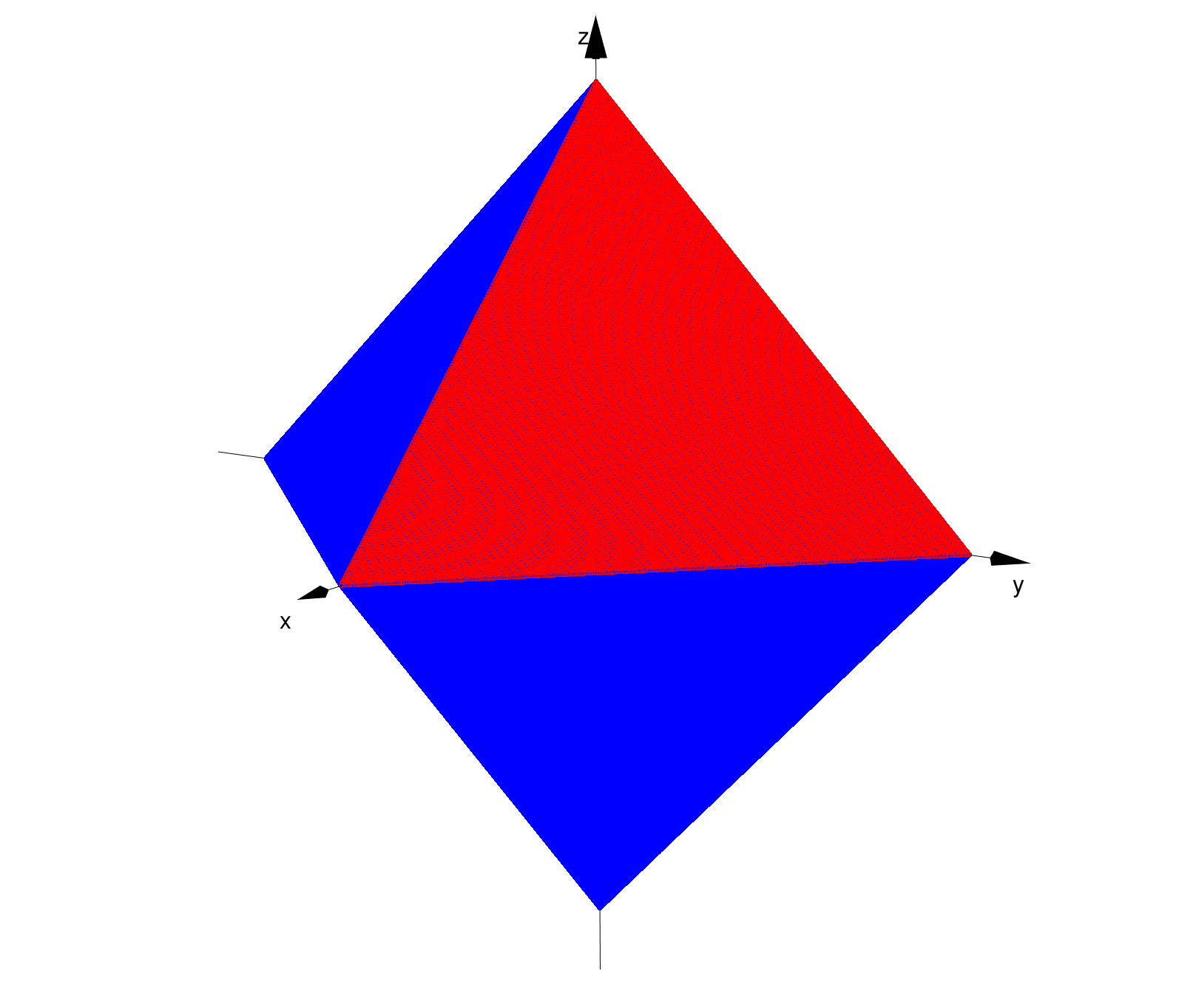}
    \end{minipage}
  %\end{narrow}
    \captionof{figure}{Two dimensional unit \(L^{p}\)-spheres with the
      corresponding \(L^{p}\) simplices for \(p = 0.25\) (left) \(p = 0.5\)
      (middle) and \(p = 1\) (right). \label{Lp.2d.spheres.p1:fig}}
\end{center}

\begin{center}
  %\begin{narrow}{-0.6in}{0in}
    \begin{minipage}{0.33\linewidth}
      \includegraphics[scale=0.11]{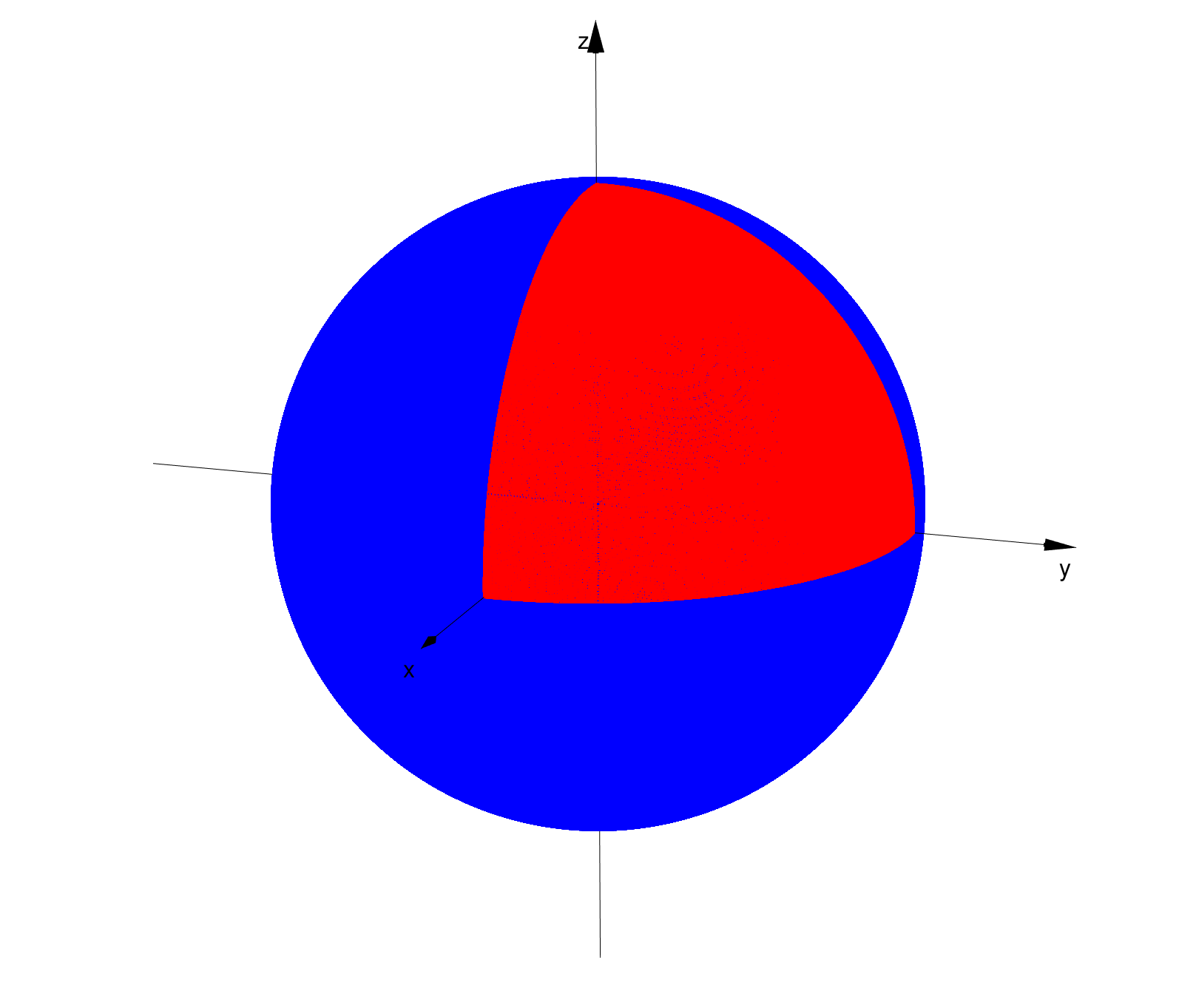}
    \end{minipage}%
    \hspace{0em}% Negative space here
    \begin{minipage}{0.33\linewidth}
      \includegraphics[scale=0.11]{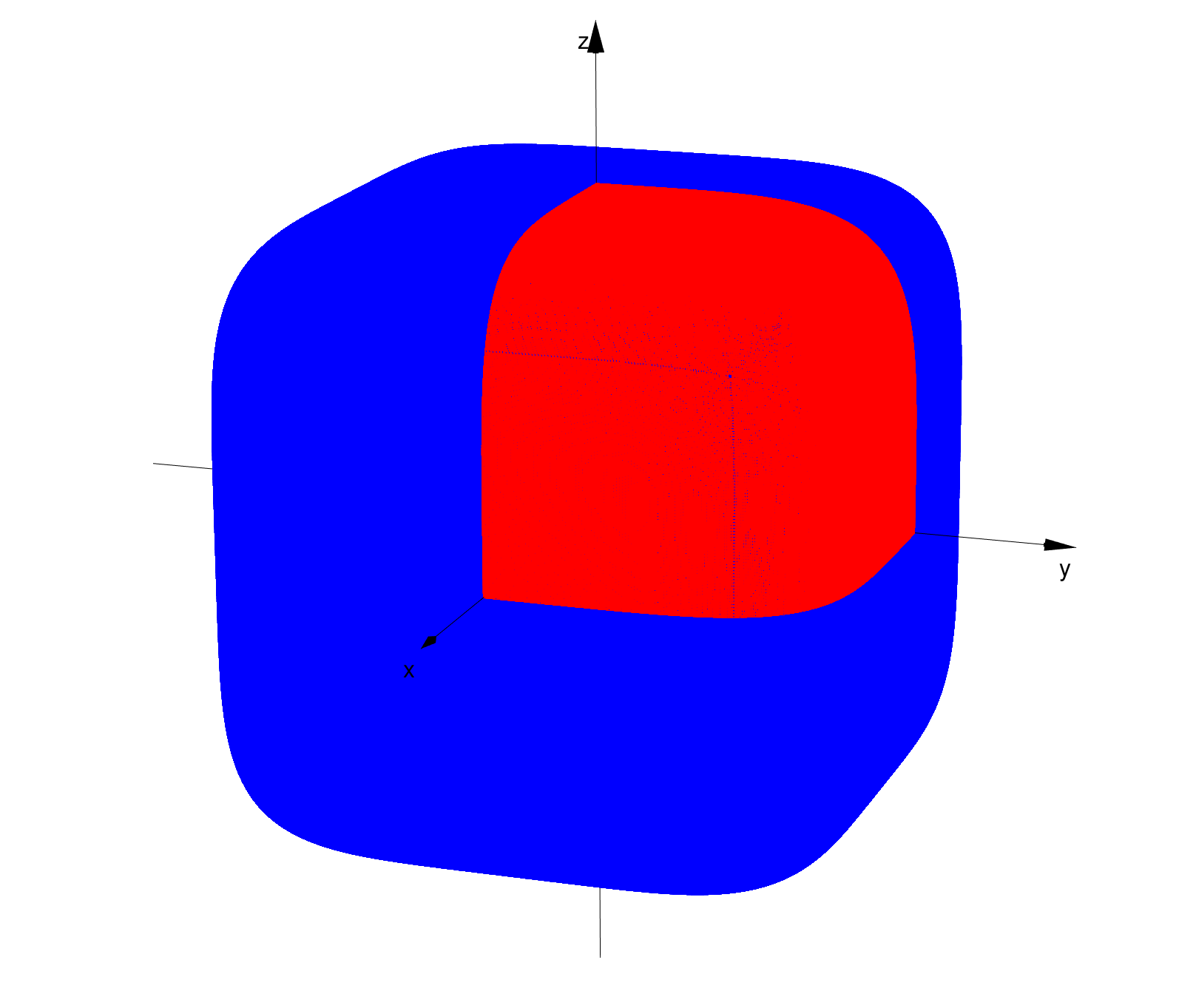}
    \end{minipage}%
    \hspace{0em}% Negative space here
    \begin{minipage}{.33\linewidth}
      \includegraphics[scale=0.11]{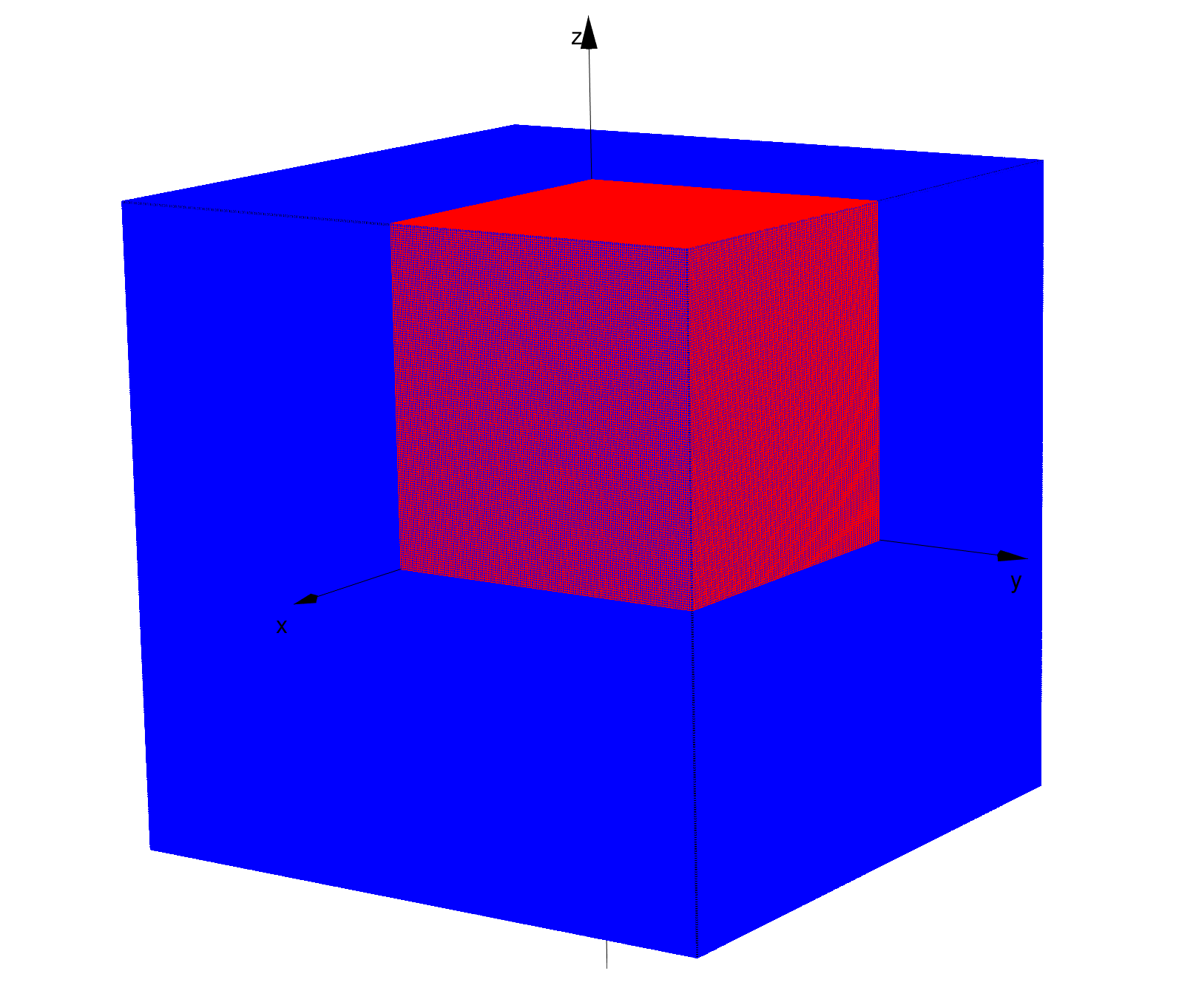}
    \end{minipage}
  %\end{narrow}
    \captionof{figure}{Two dimensional unit \(L^{p}\)-spheres with the
      corresponding \(L^{p}\) simplices for \(p = 2\) (left) \(p = 5\) (middle)
      and \(p = \infty\) (right). \label{Lp.2d.spheres.p2:fig}}
\end{center}

Figure~\ref{L1Linf.norm:fig} illustrates the process of \(L^{1}\) and
\(L^{\infty}\)-normalization on the compositional data shown in
Figure~\ref{comp.2d.ex1:fig} of Section~2.
\begin{center}
  %\begin{narrow}{-0.3in}{0in}
    \begin{minipage}[t]{.49\linewidth}
      \includegraphics[scale=0.22]{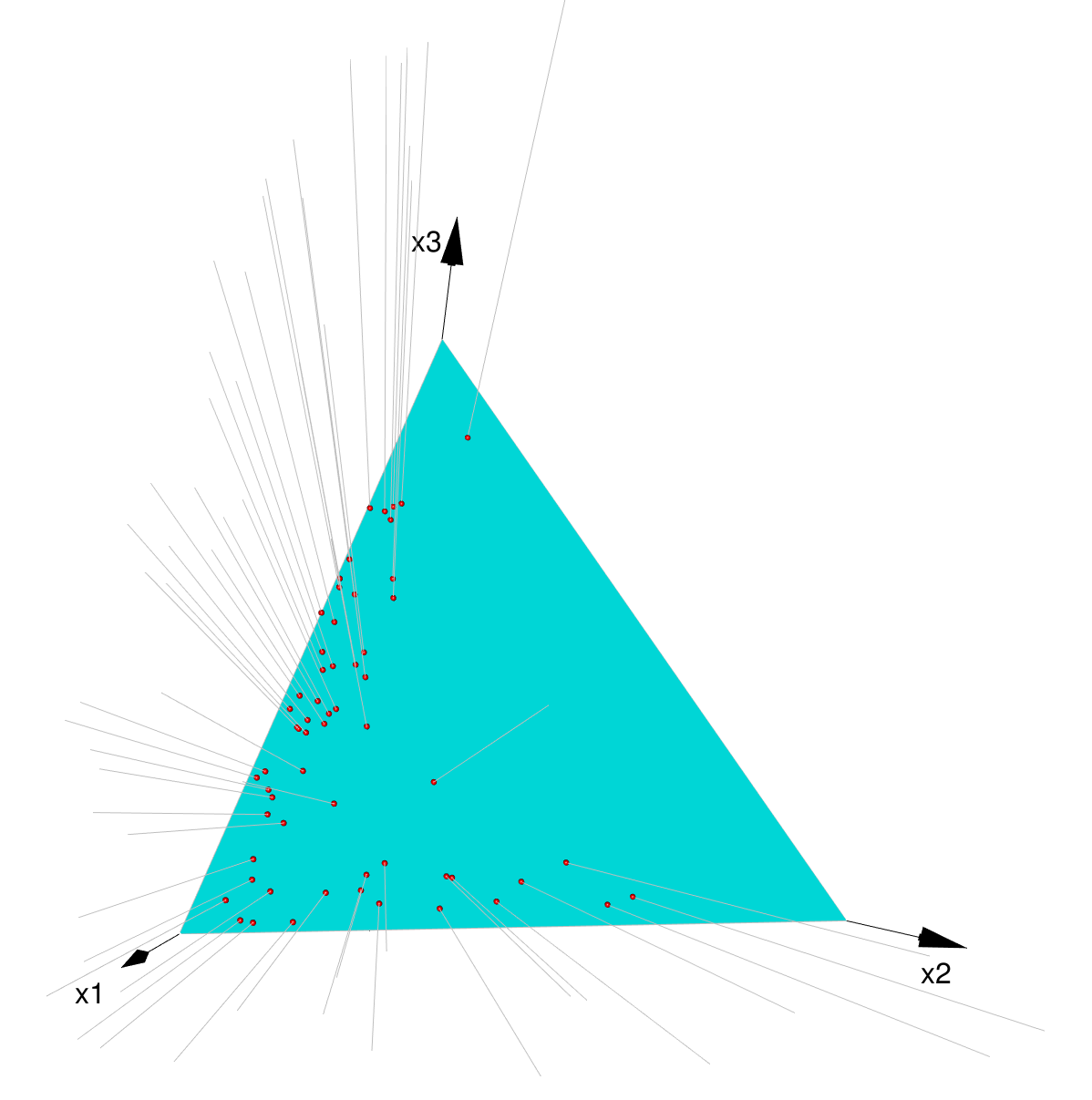}
    \end{minipage}\hspace{0cm}
    \begin{minipage}[t]{.49\linewidth}
      \includegraphics[scale=0.22]{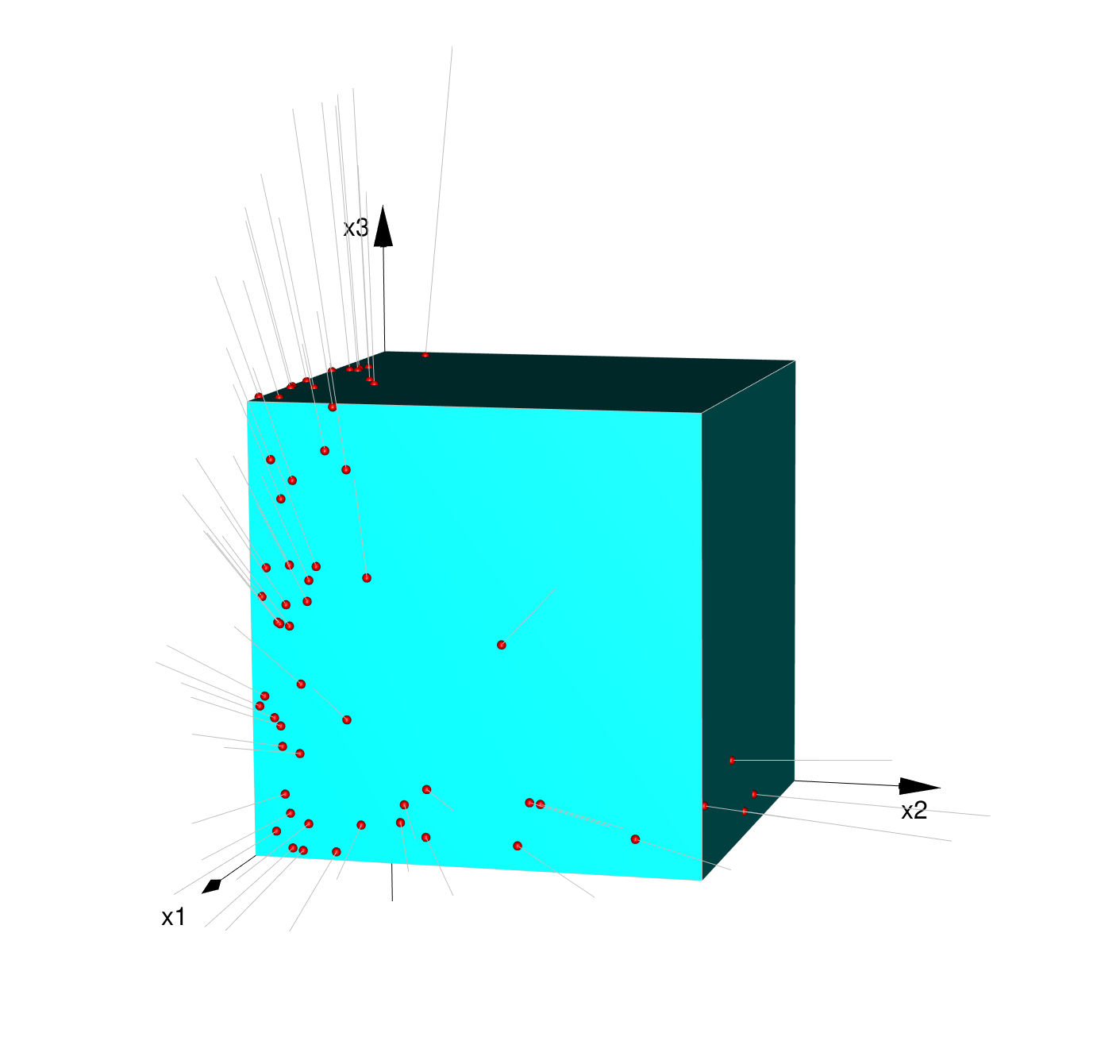}
    \end{minipage}
  %\end{narrow}
    \captionof{figure}{\(L^{1}\) and \(L^{\infty}\)-normalizations of the data
      from from Figure~\ref{comp.2d.ex1:fig} of Section~2 are the projections of
      these points, represented as gray lines, onto the standard simplex
      \(\Delta^2\) for the \(L^{1}\)-normalization (left) the
      \(L^{\infty}\)-simplex \(\Delta^2_{\infty}\) for the
      \(L^{\infty}\)-normalization (right).\label{L1Linf.norm:fig}}
\end{center}

The \(L^{\infty}\)-normalization \(\pi_{\infty}: \R^{d+1}_{\ge 0} - \{0\} \rightarrow \Delta^{d}_{\infty}\)
\[
\pi_{\infty}(x_{0}, x_{1}, \ldots, x_{d}) = \frac{1}{\| x\|_{\infty}}(x_{0}, x_{1}, \ldots, x_{d})
\]
is a single-component ratio transformation
\[
  (x_{0}, x_{1}, \ldots, x_{d}) \mapsto (\frac{x_{0}}{x_{\text{ref}}}, \frac{x_{1}}{x_{\text{ref}}}, \ldots, \frac{x_{d}}{x_{\text{ref}}})
\]
where \(x_{\text{ref}}\) is the maximum component value \( \| x\|_{\infty} \).
This implies that \(\pi_{\infty}\) is subcompositionally coherent with respect
to all sub-setting of component operations involving components that are not
maximal in any sample.

The special role of \(L^{\infty}\)-normalization is highlighted by Greenacre's
finding, which shows that PCA analysis on data transformed using centered log
ratio
\[
  \pi_{\text{CLR}}(x_{0}, x_{1}, \ldots, x_{d}) = \left(\log\left(\frac{x_{0}}{g(x)}\right), \ldots, \log\left(\frac{x_{d}}{g(x)}\right)\right),
\]
within the interior of the standard simplex, where
\(g(x) = \big( \prod_{i=0}^{d}x_{i} \big)^{1/(d+1)}\) is the geometric mean
of \(x = (x_{0}, x_{1}, \ldots, x_{d})\), can be viewed as the limit of
correspondence analysis (CA) on \(L^{1}\)-normalized data that has undergone
power transformation and rescaling \cite{greenacre2009power}. The power
transformation \(r_{p}: \Delta^{d} \rightarrow \Delta^{d}_{p}\)
\[
  r_{p}(x_{0}, x_{1}, \ldots, x_{d}) = (x_{0}^{1/p}, x_{1}^{1/p}, \ldots, x_{d}^{1/p})
\]
is a homeomorphism between the standard simplex \(\Delta^{d}\) and the
\(L^{p}\)-simplex \(\Delta^{d}_{p}\) for any \(p \ne 2\). When compositional
data are power transformed with \(r_{p}\), re-scaled row-wise, and analyzed with
CA, followed by rescaling the solution by \(p\), this process converges to PCA
on the CLR transformed data as \(p\) approaches \(\infty\). Geometrically, as \(p\)
approaches \(\infty\), the \(L^{p}\)-simplex \(\Delta^{d}_{p}\) converges to
the \(L^{\infty}\)-simplex \(\Delta^{d}_{\infty}\) and in the limit the
\(L^{p}\) normalization becomes \(L^{\infty}\) normalization.

%%% Local Variables:
%%% mode: latex
%%% TeX-master: "Linf_paper"
%%% End:
 % 4
\section*{5.  \(L^{\infty}\)-Decomposition of Compositional Spaces}

The \(L^{\infty}\)-simplex \(\Delta_{\infty}^{d}\) has a
\(L^{\infty}\)-decomposition into \((d+1)\) \(d\)-dimensional
\(L^{\infty}\)-cells
\[
  \Delta_{\infty}^{d} = Q_{0} \cup Q_{1} \cup \ldots \cup Q_{d}
\]
with \(Q_{k}\) defined as the intersection of the \(L^{\infty}\)-simplex
\(\Delta_{\infty}^{d}\) with the hyper-plane \(x_{k} = 1\). Thus, \(Q_{k}\),
consists of samples where the absolute abundance of the \(k\)-th component is
equal to or greater than the absolute abundances of all other components.
\(Q_{k}\) is homeomorphic with the unit hypercube \([0,1]^{d}\) as the condition
\(x_{k} = 1\) implies that the other than \(x_{k}\) coordinates of the
\(L^{\infty}\)-simplex can take any value in the closed interval \([0,1]\).
Since there are \(d\) such coordinates (after excluding \(x_{k}\)) \(Q_{k}\) can
be identified with the unit hypercube \([0,1]^{d}\).

Since \(\mathbb{RP}^{d}_{\ge 0}\) is homeomorphic to \(\Delta_{\infty}^{d}\),
the \(L^{\infty}\)-decomposition of \(\Delta_{\infty}^{d}\) into top dimensional
cells induces an \(L^{\infty}\)-decomposition of \(\mathbb{RP}^{d}_{\ge 0}\)
\[
  \mathbb{RP}^{d}_{\ge 0} = Q_{0}^{c} \cup Q_{1}^{c} \cup \ldots \cup Q_{d}^{c}
\]
where \(Q_{k}^{c} \) is the set of lines through the origin in
\(\R^{d+1}_{\ge 0}\) that pass through \(Q_{k}\). Thus, \(Q_{k}^{c} \) consists
of compositions \([x_{0}: x_{1}: \ldots : x_{d}]\), such that
\(x_{k} \ge x_{i}\) for all \(i \ne k\). The \(L^{\infty}\)-decomposition of
\(\mathbb{RP}^{d}_{\ge 0}\) induces a decomposition into \(d+1\) components of
any parametrization of \(\mathbb{RP}^{d}_{\ge 0}\). In particular, the standard
simplex, \(\Delta^{d}\), has an \(L^{\infty}\)-decomposition as illustrated on
the following figure.
\begin{center}
\includegraphics[scale=0.6]{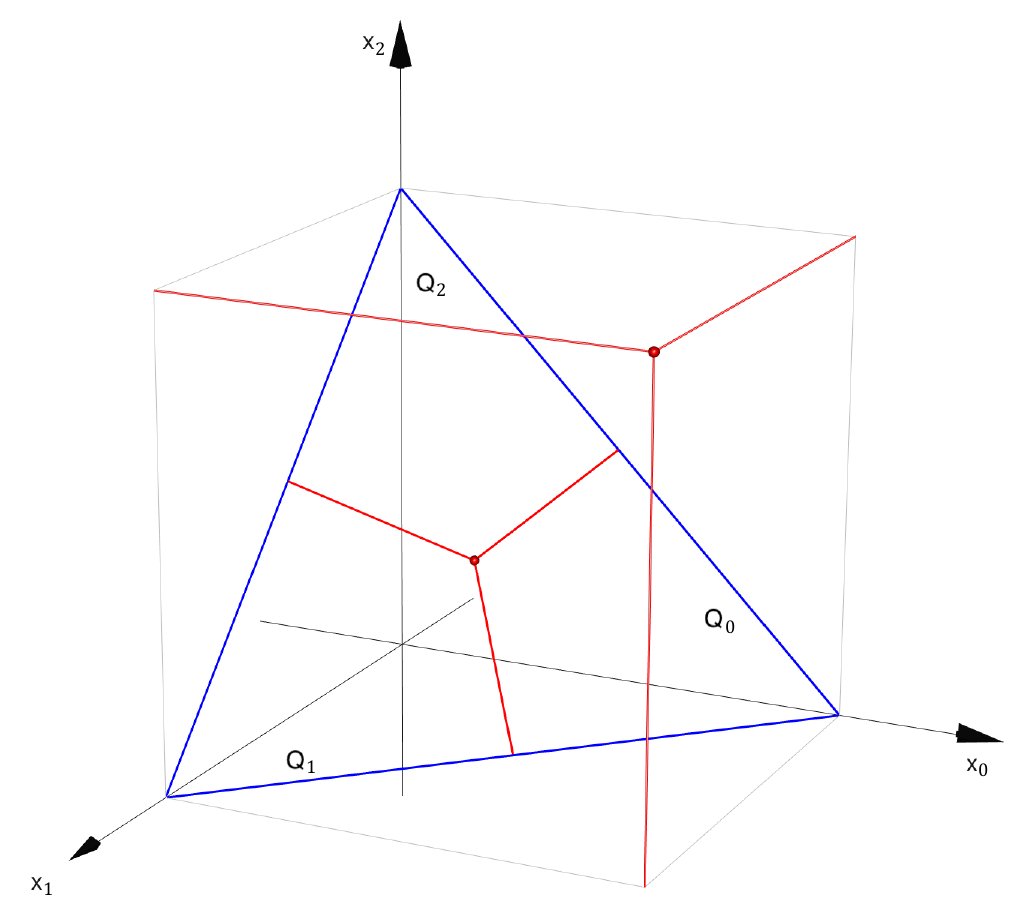}
\captionof{figure}{\(L^{\infty}\)-decomposition of the \(L^{\infty}\) and
  standard two-dimensional simplices. \label{LinfDelta.onto.L1Delta.2d:fig}}
\end{center}

\begin{center}
\includegraphics[scale=0.6]{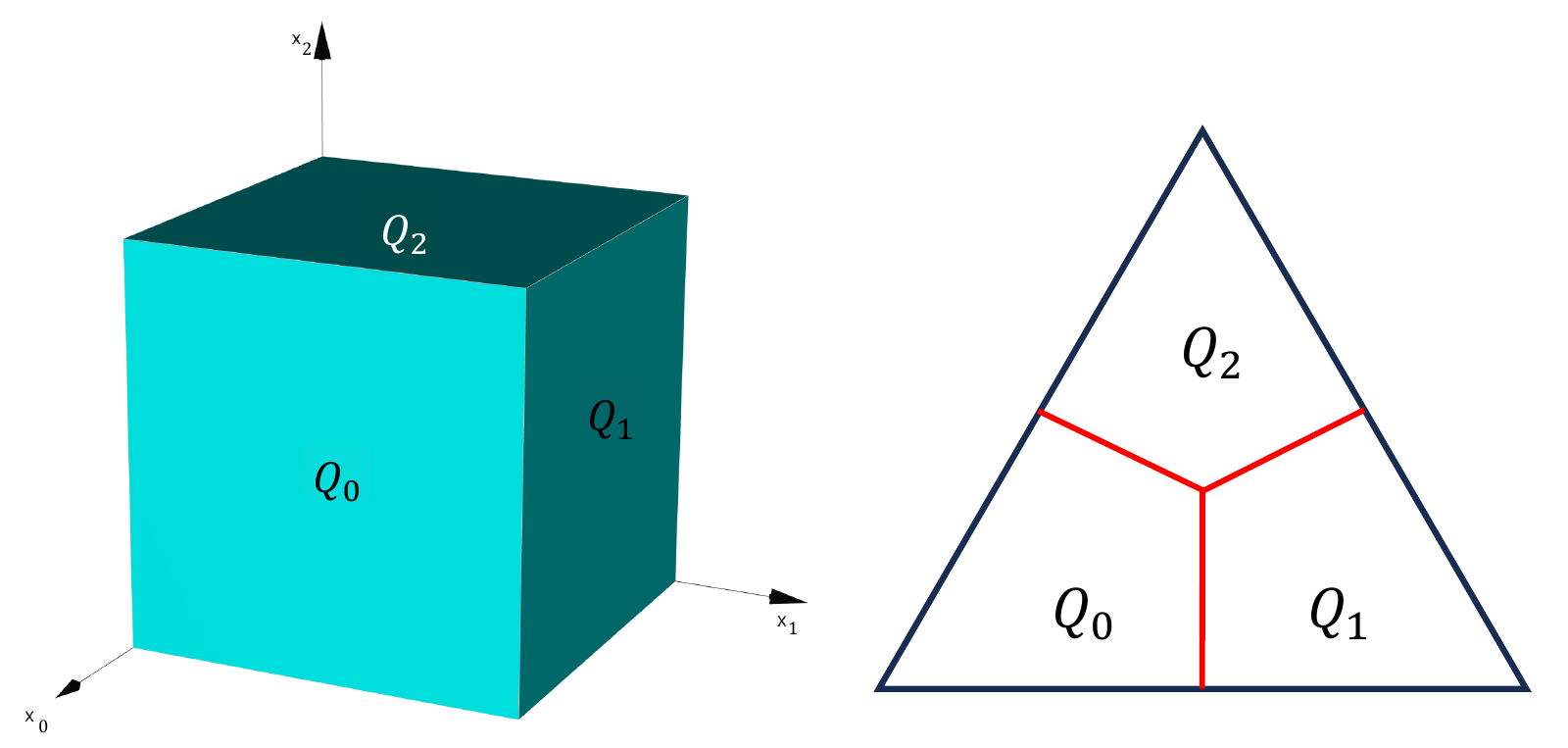}
\captionof{figure}{\(L^{\infty}\)-decomposition of the \(L^{\infty}\) and
  standard three-dimensional simplices.  \label{LinfDelta.onto.L1Delta.2d.p2:fig}}
\end{center}
\(L^{\infty}\)-decomposition of \(\mathbb{RP}^{d}_{\ge 0}\) allows systematic
study of the local structure of compositional data, by restricting attention to
the subsets of data contained in different \(Q^{c}_{k}\) region of
\(\mathbb{RP}^{d}_{\ge 0}\). One may be concerned by the fact that with large
number of components, a number that may be in thousands, it may be impractical
to perform analyses on large number of \(Q^{c}_{k}\) cells. Typically, this is
not the case as shown in Table~\ref{Linf.top.cell.freq.asv.CMB:tbl} where 96.9\%
of 13,231 samples of the combined vaginal 16S rRNA amplicon data
\cite{france:2020} is contained in the first 13 \(L^{\infty}\)-cells with the
first two cells containing almost 60\% of all data. The number of components in
this dataset is 199.
\begin{center}
\begin{tabular}{lrrrrr}
\hline\hline
\multicolumn{1}{l}{Phylotype}&\multicolumn{1}{r}{Freq}&\multicolumn{1}{r}{Perc}&\multicolumn{1}{r}{CumPerc}&\multicolumn{1}{r}{n(det)}&\multicolumn{1}{r}{p(det)}\tabularnewline
\hline
\emph{Lactobacillus iners}&4068&30.7&30.7&11884&89.8\tabularnewline
\emph{Lactobacillus crispatus}&3686&27.9&58.6&10605&80.2\tabularnewline
\emph{Gardnerella vaginalis}&2422&18.3&76.9&10272&77.6\tabularnewline
\emph{BVAB1}&657&5.0&81.9&5446&41.2\tabularnewline
\emph{Lactobacillus gasseri}&428&3.2&85.1&5980&45.2\tabularnewline
\emph{Lactobacillus jensenii}&414&3.1&88.2&6396&48.3\tabularnewline
\emph{Atopobium vaginae}&280&2.1&90.4&7520&56.8\tabularnewline
\emph{g Streptococcus}&250&1.9&92.2&8128&61.4\tabularnewline
\emph{Sneathia sanguinegens}&236&1.8&94.0&6067&45.9\tabularnewline
\emph{g Bifidobacterium}&178&1.3&95.4&3166&23.9\tabularnewline
\emph{g Enterococcus}&68&0.5&95.9&2536&19.2\tabularnewline
\emph{g Anaerococcus}&67&0.5&96.4&10085&76.2\tabularnewline
\emph{g Corynebacterium} 1&64&0.5&96.9&8677&65.6\tabularnewline
\hline
\end{tabular}
\captionof{table}{Frequencies, percentages and cumulative percentages of
  samples of the \(L^{\infty}\)-cells as well as the number and percentage of
  samples where the corresponding phylotype was detected in the combined vaginal
  16S rRNA amplicon data
  \cite{france:2020}.\label{Linf.top.cell.freq.asv.CMB:tbl}}
\end{center}

Typically, points found in \(L^{\infty}\)-cells with few samples tend to be
positioned at the boundaries near cells with a markedly larger sample size. This
motivates that following construct of a truncated \(L^{\infty}\)-decomposition
of a compositional dataset. Let \(n_{0}\) be the minimal number of samples each
\(L^{\infty}\)-cell is required to have. An \(n_{0}\)-truncated
\(L^{\infty}\)-decomposition of the data is constructed from the
\(L^{\infty}\)-cells that contain at least \(n_{0}\) elements in the following
fashion. By reordering the components if necessary, we assume the first \(m\)
\(L^{\infty}\)-cells, \(Q_{0}, Q_{1}, \ldots, Q_{m}\), contain at least
\(n_{0}\) elements. Any point \([x]\) from an \(L^{\infty}\)-cell \(Q_{k}\) with
\(k > m\) is reassigned to the cell \(Q_{i}\) if the \(i\)-th component of
\([x]\) has the highest value among the first \(m\) components of \([x]\).

\(L^{\infty}\)-decomposition can be further refined by subdividing each
\(L^{\infty}\)-cell into \(2^{d}\) equal size sub-hypercubes. This produces a
very large number sub-cells. Typically, a very small subset of these holds the
data. The sub-cells with small number of samples can be merged with cells
carrying more samples using the truncation algorithm outlined above.

%%% Local Variables:
%%% mode: latex
%%% TeX-master: "Linf_paper"
%%% End:
 % 5
\section*{6. CSTs and \(L^{\infty}\)-CSTs }

Community state types (CSTs) were originally defined as clusters of vaginal
samples derived using hierarchical clustering with Ward linkage over a vaginal
16S rRNA amplicon data \cite{ravel2011vaginal}. They were introduced as a tool
to facilitate a high level characterization of the structure of vaginal
microbial communities and have been instrumental in advancing our understanding
of conditions like bacterial vaginosis and preterm delivery
\cite{romero2014vaginal,tuddenham2019associations,o2020asymptomatic}. In order
to remove the dependence of CST assignment on the clustering of the data a
VALENCIA CST classifier was created \cite{france:2020}.
\begin{center}
\includegraphics[scale=0.6]{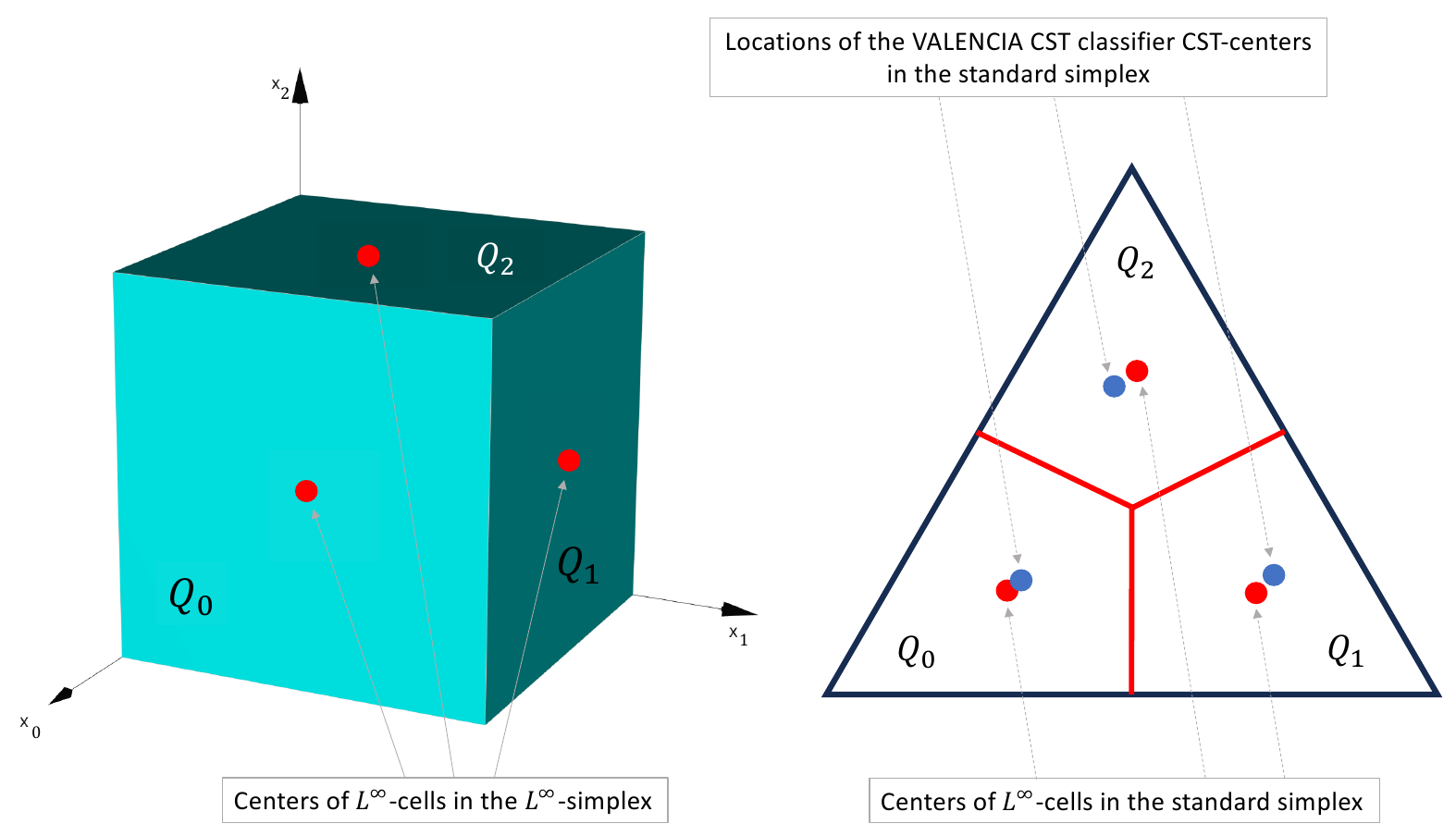}
\captionof{figure}{The mapping of \(L^{\infty}\)-cells from the
  \(L^{\infty}\)-simples (left) into the standard simplex (right) maps the
  centers of the \(L^{\infty}\)-simplex cells into the centers of the
  corresponding regions in the standard simplex (shown as red disks). VALENCIA
  CST classifier uses CST centers (shown as blue disks) in the standard simplex,
  derived from the training dataset of large number of vaginal samples, to
  classify vaginal samples to CSTs. The comparison of VALENCIA CSTs and
  \(L^{\infty}\)-CSTs tables indicate that the VALENCIA CST centers are located
  close to the centers of the corresponding \(L^{\infty}\)-cells (shown in red)
  in the standard simplex. \label{Linf.cells.and.Valencia.CST.classifier:fig}}
\end{center}

The following tables compare VALENCIA CSTs with the \(L^{\infty}\)-cells with at
least 50 samples within the dataset consisting of 13,231 vaginal samples
\cite{france:2020} showing a high level of concordance between
\(L^{\infty}\)-cells and VALENCIA CSTs. Thus, one can think of
\(L^{\infty}\)-cells as an alternative construct of CSTs. This perspective
shifts the understanding of CSTs and enterotypes from clusters to groupings of
samples based on patterns of species dominance. Since \(L^{\infty}\)-cells are
defined using ratios of relative abundances, that are the same as ratio of
absolute abundances, \(L^{\infty}\)-cells define groups of samples where the
absolute abundance of one species (the reference of the given cell) is at least
as high or higher than the absolute abundance of other species within the given
set of samples. This approach provides clear criteria for assigning samples to
CSTs or enterotypes, addressing the following limitations inherent in
clustering-based approaches for constructing CSTs or enterotypes.

1) The definition of CSTs and enterotypes is highly dependent on the choice of a
clustering algorithm. This has led to a situation in which distinct research
groups adopt different algorithms for defining CSTs or enterotypes, making it
impossible to compare their results with the results of other groups based on
different definition of CSTs or enterotypes.

2) Since CSTs and enterotypes are defined using complex clustering algorithms, they
don't offer clear, easily understandable criteria that determine why a sample is
for example classified to CST I, effectively acting as a black box classifier.

3) The structure of the space of microbial community states is a continuum,
rather than a collection of discrete, distinct clusters. Consequently, the
application of clustering strategies that artificially segment this continuum is
difficult to justify.

4) Clustering-based construction of CSTs and enterotypes is dataset dependent,
whereas \(L^{\infty}\)-cell assignemt is sample-dependant. Altering the dataset
by adding or removing samples can shift the assignment of CSTs or enterotypes.
On the other hand, an assignemt to a \(L^{\infty}\)-cell does not depend on
other samples. \(L^{\infty}\)-CST assignment depends on other samples only for a
minotiry of rare samples. One solution to this CST/enterotype problem is to
develop a classifier trained on the entirety of the available data
\cite{france:2020}. However, this approach introduces a new challenge: as new
data emerges, the algorithm must be updated, resulting in revised CSTs that may
not align with previous versions.

The mentioned challenges significantly complicate the creation of CSTs in new
microbiome contexts, where identifying meaningful clusters is inherently
difficult due to the absence of clear groupings.

The use of \(L^{\infty}\)-cells offers a significant advantage for high-level
characterization of microbiota. These cells are not just groupings of samples;
they also incorporate a homogeneous coordinate system that enhances the
elucidation of their internal structure. As discussed in Section~8, this
coordinate system can be extended to include all samples, even those where the
denominator is zero. This extension effectively maps each \(L^{\infty}\)-cell to
a subset of a hypercube \([0,1]^{d}\). This mapping facilitates a comprehensive
study of the global structure of microbial community states from various
perspectives.
\begin{center}
\begin{tabular}{lrrrrrrr}
\hline\hline
\multicolumn{1}{l}{\(L^{\infty}\)-cell}&\multicolumn{1}{r}{I}&\multicolumn{1}{r}{II}&\multicolumn{1}{r}{III}&\multicolumn{1}{r}{IV-A}&\multicolumn{1}{r}{IV-B}&\multicolumn{1}{r}{IV-C}&\multicolumn{1}{r}{V}\tabularnewline
\hline
\emph{Lactobacillus iners}&0.0&0.0&\textbf{93.1}&0.7&3.6&0.3&2.3\tabularnewline
\emph{Lactobacillus crispatus}&\textbf{98.6}&0.0&1.0&0.0&0.1&0.2&0.0\tabularnewline
\emph{Gardnerella vaginalis}&0.2&0.7&0.4&5.9&\textbf{92.5}&0.1&0.3\tabularnewline
\emph{BVAB1}&0.0&0.0&0.0&\textbf{100.0}&0.0&0.0&0.0\tabularnewline
\emph{Lactobacillus gasseri}&0.2&\textbf{99.5}&0.0&0.0&0.2&0.0&0.0\tabularnewline
\emph{Lactobacillus jensenii}&0.5&0.0&0.0&0.0&0.0&0.5&\textbf{99.0}\tabularnewline
\emph{Atopobium vaginae}&0.4&0.0&0.7&3.2&\textbf{92.5}&0.4&2.9\tabularnewline
\emph{g Streptococcus}&0.0&0.0&0.0&0.0&0.4&\textbf{99.6}&0.0\tabularnewline
\emph{Sneathia sanguinegens}&0.0&0.0&1.3&15.7&\textbf{80.9}&1.7&0.4\tabularnewline
\emph{g Bifidobacterium}&0.0&0.0&0.0&0.0&1.1&\textbf{98.9}&0.0\tabularnewline
\emph{g Enterococcus}&0.0&0.0&0.0&0.0&0.0&\textbf{100.0}&0.0\tabularnewline
\emph{g Anaerococcus}&1.5&0.0&0.0&1.5&23.9&\textbf{73.1}&0.0\tabularnewline
\emph{g Corynebacterium} 1&1.6&3.1&0.0&0.0&1.6&\textbf{93.8}&0.0\tabularnewline
\hline
\end{tabular}
\captionof{table}{Percentages of \(L^{\infty}\)-cells present in different CSTs
  using Valencia 16S rRNA data. Only \(L^{\infty}\)-cells with at least 50
  samples were analyzed.\label{Linf.cells.vs.CSTs:tbl}}
\end{center}

\begin{center}
  {\footnotesize
\begin{tabular}{lrrrrrrrrrrrrr}
\hline\hline
\multicolumn{1}{l}{\(L^{\infty}\)-cell}&\multicolumn{1}{r}{I-A}&\multicolumn{1}{r}{I-B}&\multicolumn{1}{r}{II}&\multicolumn{1}{r}{III-A}&\multicolumn{1}{r}{III-B}&\multicolumn{1}{r}{IV-A}&\multicolumn{1}{r}{IV-B}&\multicolumn{1}{r}{IV-C0}&\multicolumn{1}{r}{IV-C1}&\multicolumn{1}{r}{IV-C2}&\multicolumn{1}{r}{IV-C3}&\multicolumn{1}{r}{IV-C4}&\multicolumn{1}{r}{V}\tabularnewline
\hline
\emph{Lactobacillus iners}&0.0&0.0&0.0&56.2&36.9&0.7&3.6&0.2&0.0&0.0&0.0&0.1&2.3\tabularnewline
\emph{Lactobacillus crispatus}&68.4&30.2&0.0&0.0&1.0&0.0&0.1&0.2&0.0&0.0&0.0&0.0&0.0\tabularnewline
\emph{Gardnerella vaginalis}&0.0&0.2&0.7&0.0&0.4&5.9&92.5&0.0&0.1&0.0&0.0&0.0&0.3\tabularnewline
\emph{BVAB1}&0.0&0.0&0.0&0.0&0.0&100.0&0.0&0.0&0.0&0.0&0.0&0.0&0.0\tabularnewline
\emph{Lactobacillus gasseri}&0.0&0.2&99.5&0.0&0.0&0.0&0.2&0.0&0.0&0.0&0.0&0.0&0.0\tabularnewline
\emph{Lactobacillus jensenii}&0.0&0.5&0.0&0.0&0.0&0.0&0.0&0.5&0.0&0.0&0.0&0.0&99.0\tabularnewline
\emph{Atopobium vaginae}&0.0&0.4&0.0&0.0&0.7&3.2&92.5&0.0&0.4&0.0&0.0&0.0&2.9\tabularnewline
\emph{g Streptococcus}&0.0&0.0&0.0&0.0&0.0&0.0&0.4&3.2&95.2&0.4&0.8&0.0&0.0\tabularnewline
\emph{Sneathia sanguinegens}&0.0&0.0&0.0&0.0&1.3&15.7&80.9&0.8&0.4&0.4&0.0&0.0&0.4\tabularnewline
\emph{g Bifidobacterium}&0.0&0.0&0.0&0.0&0.0&0.0&1.1&0.0&0.0&0.0&98.9&0.0&0.0\tabularnewline
\emph{g Enterococcus}&0.0&0.0&0.0&0.0&0.0&0.0&0.0&0.0&0.0&100.0&0.0&0.0&0.0\tabularnewline
\emph{g Anaerococcus}&0.0&1.5&0.0&0.0&0.0&1.5&23.9&71.6&1.5&0.0&0.0&0.0&0.0\tabularnewline
\emph{g Corynebacterium} 1&0.0&1.6&3.1&0.0&0.0&0.0&1.6&87.5&1.6&3.1&1.6&0.0&0.0\tabularnewline
\hline
\end{tabular}}
\captionof{table}{Percentage of \(L^{\infty}\)-cells present in different
  sub-CSTs using Valencia 16S rRNA data. Only \(L^{\infty}\)-cells with at least
  50 samples were analyzed.\label{Linf.cells.vs.SUBCSTs:tbl}}
\end{center}

\begin{center}
\begin{tabular}{lrrrrrrr}
\hline\hline
\multicolumn{1}{l}{\(L^{\infty}\)-cell}&\multicolumn{1}{r}{III}&\multicolumn{1}{r}{I}&\multicolumn{1}{r}{IV-B}&\multicolumn{1}{r}{IV-A}&\multicolumn{1}{r}{IV-C}&\multicolumn{1}{r}{V}&\multicolumn{1}{r}{II}\tabularnewline
\hline
\emph{Atopobium vaginae}&0.1&0.0&9.1&1.0&0.2&1.5&0.0\tabularnewline
\emph{BVAB1}&0.0&0.0&0.0&74.9&0.0&0.0&0.0\tabularnewline
\emph{g Anaerococcus}&0.0&0.0&0.6&0.1&7.8&0.0&0.0\tabularnewline
\emph{g Bifidobacterium}&0.0&0.0&0.1&0.0&27.9&0.0&0.0\tabularnewline
\emph{g Corynebacterium} 1&0.0&0.0&0.0&0.0&9.5&0.0&0.4\tabularnewline
\emph{g Enterococcus}&0.0&0.0&0.0&0.0&10.8&0.0&0.0\tabularnewline
\emph{g Streptococcus}&0.0&0.0&0.0&0.0&39.5&0.0&0.0\tabularnewline
\emph{Gardnerella vaginalis}&0.2&0.1&78.4&16.3&0.3&1.5&3.6\tabularnewline
\emph{Lactobacillus crispatus}&1.0&99.7&0.1&0.1&1.4&0.2&0.0\tabularnewline
\emph{Lactobacillus gasseri}&0.0&0.0&0.0&0.0&0.0&0.0&95.5\tabularnewline
\emph{Lactobacillus iners}&98.6&0.0&5.1&3.3&1.7&17.7&0.4\tabularnewline
\emph{Lactobacillus jensenii}&0.0&0.1&0.0&0.0&0.3&78.8&0.0\tabularnewline
\emph{Sneathia sanguinegens}&0.1&0.0&6.7&4.2&0.6&0.2&0.0\tabularnewline
\hline
\end{tabular}
\captionof{table}{Percentage of CSTs present in different \(L^{\infty}\)-cells
  using Valencia 16S rRNA data. Only \(L^{\infty}\)-cells with at least 50
  samples were analyzed.\label{Linf.cells.vs.CSTs:tbl2}}\end{center}

Based on the notable agreement between \(L^{\infty}\)-cells and CSTs, we define
the cells of the truncated \(L^{\infty}\)-decomposition, with the minimal cell
size of 50 samples, as \(L^{\infty}\)-CSTs. In the truncated
\(L^{\infty}\)-decomposition samples from \(L^{\infty}\)-cells with less than 50
samples are reassigned to those with at least 50 samples. This rearrangement is
supported by the observation that samples in lesser-populated
\(L^{\infty}\)-cells are typically situated near the boundary of that cell
adjoining \(L^{\infty}\)-cells with a high sample count.

The mapping between \(L^{\infty}\)-CSTs and VALENCIA CSTs is presented in the
Table~\ref{Linf.CSTs.to.Valencia.CSTs:tbl}. Each \(L^{\infty}\)-CST is assigned
with the VALENCIA CST with the largest number of samples within that
\(L^{\infty}\)-CST. It is important to note that none of the \(L^{\infty}\)-CSTs
was assigned to CST I-B due to the predominant presence of CST I-B samples in
the \emph{Lactobacillus crispatus} \(L^{\infty}\)-CST, with the remainder
scattered across \(L^{\infty}\)-cells holding less than 50 samples each.
Additionally, three \(L^{\infty}\)-CSTs correspond to CST IV-B and two to CST
IV-C0, thereby segmenting these CSTs into distinct groups based on their unique
absolute abundance patterns.
\begin{center}
\begin{tabular}{lrrrr}
\hline\hline
\multicolumn{1}{l}{\(L^{\infty}\)-CST}&\multicolumn{1}{r}{CST}&\multicolumn{1}{r}{n(comm)}&\multicolumn{1}{r}{n(CST)}&\multicolumn{1}{r}{n(\(L^{\infty}\)-CST)}\tabularnewline
\hline
\emph{Lactobacillus crispatus}&I-A&2522&2522&3702\tabularnewline
\emph{Lactobacillus gasseri}&II&432&454&436\tabularnewline
\emph{Lactobacillus iners}&III-A&2288&2288&4116\tabularnewline
\emph{BVAB1}&IV-A&679&925&679\tabularnewline
\emph{Atopobium vaginae}&IV-B&283&3018&312\tabularnewline
\emph{Gardnerella vaginalis}&IV-B&2340&3018&2549\tabularnewline
\emph{Sneathia sanguinegens}&IV-B&212&3018&265\tabularnewline
\emph{g Anaerococcus}&IV-C0&80&210&103\tabularnewline
\emph{g Corynebacterium} 1&IV-C0&71&210&85\tabularnewline
\emph{g Streptococcus}&IV-C1&254&260&279\tabularnewline
\emph{g Enterococcus}&IV-C2&89&95&96\tabularnewline
\emph{g Bifidobacterium}&IV-C3&185&189&192\tabularnewline
\emph{Lactobacillus jensenii}&V&412&522&417\tabularnewline
\hline
\end{tabular}
\captionof{table}{Mapping of \(L^{\infty}\)-CSTs onto VALENCIA CSTs with the
  number of samples common to the corresponding CSTs (column n(comm)), the
  number of samples in the given VALENCIA CST (column n(CST)) and the number of
  samples in the corresponding \(L^{\infty}\)-CST (column
  \(L^{\infty}\)-CST). \label{Linf.CSTs.to.Valencia.CSTs:tbl}}
\end{center}

Given the reduced prevalence of single-species dominance within the gut
microbiome, it may be important to examine patterns where a consortium of
bacteria collectively dominates a community. This shift in perspective can
unveil more profound insights into the structure gut microbiome's community
state space. Consequently, this analysis requires transitioning from focusing on
top-dimensional \(L^{\infty}\)-cells to exploring lower-dimensional
\(L^{\infty}\)-cells defined as the intersection of several top-dimensional
\(L^{\infty}\)-cells, offering a more nuanced view of microbial dominance.

%%% Local Variables:
%%% mode: latex
%%% TeX-master: "Linf_paper"
%%% End:
 % 6
\section*{7. Alignment of \(L^{\infty}\)-cells through rotation }

One-dimensional \(L^{\infty}\)-simplex \(\Delta_{\infty}^{1}\) decomposes as a
union of two unit intervals \(Q_0 = \{1\} \times [0,1]\) and
\(Q_1 = [0,1] \times \{1\}\) (see the left panel of
Figure~\ref{DLinf1.rotation:fig}). If we rotate the \(L^{\infty}\)-cell \(Q_0\)
around the point \((1,1)\), keeping the vertex \((1,1)\) of \(Q_0\) fixed, to
the horizontal position, the image \(rQ_0\) of \(Q_{0}\) after rotation will be
aligned with \(Q_1\), both lying on the \(x_1 = 1\) line, with \(rQ_0\)
following \(Q_1\) (see the right panel of Figure~\ref{DLinf1.rotation:fig}). The
new coordinates on the rotated \(Q_0\) are \((2-y,1)\), where \(y \in [0,1]\)
and \(Q_0 = \{(1,y): y\in [0,1]\}\). Thus, the rotation of \(Q_0\) is given by
the mapping. \((1,y) \mapsto (2 - y, 1)\). This, gives an explicit homeomorphism
between \(\Delta_{\infty}^{1}\) and the interval \([0,2]\).
\begin{center}
  \includegraphics[scale=0.5]{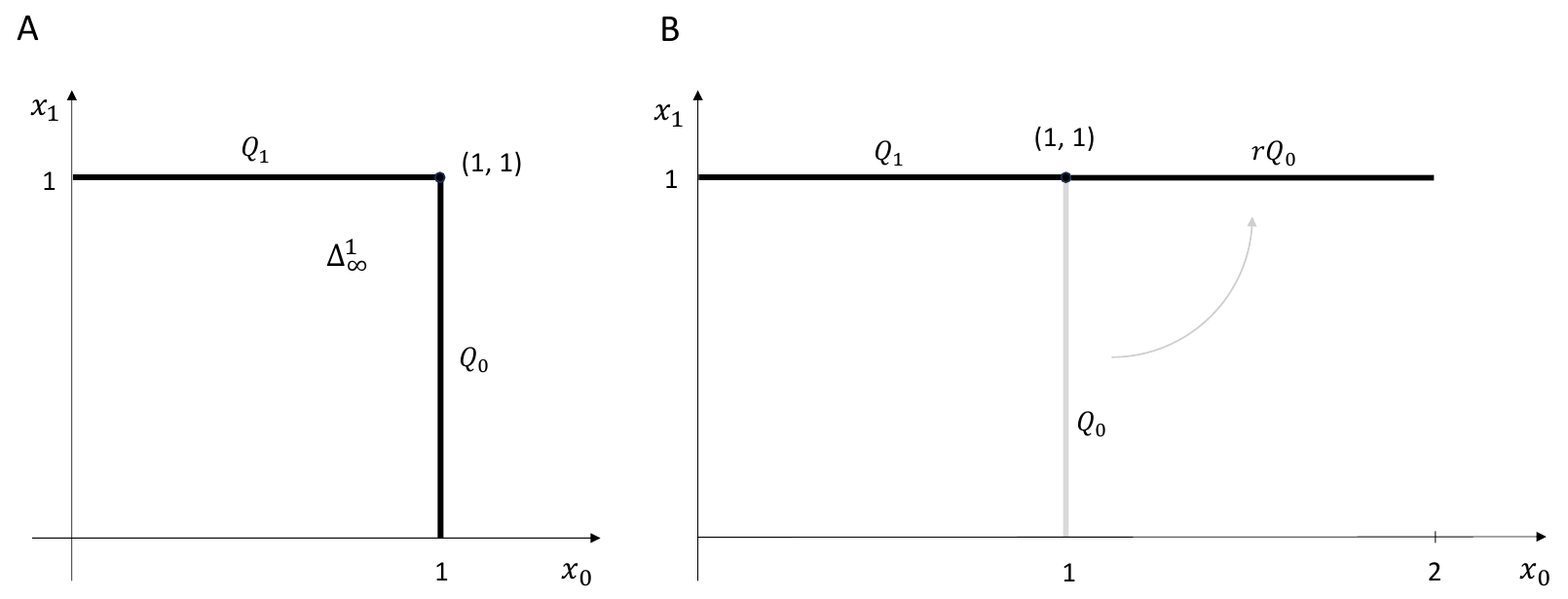}
  \captionof{figure}{Rotation of the face \(Q_0\) around the point
    \((1,1)\). \label{DLinf1.rotation:fig}}
\end{center}
The same rotation procedure can be applied to \(L^{\infty}\)-cells of the
\(L^{\infty}\)-simplex in any dimension as shown in the two-dimensional case in
the following figure.
\begin{center}
  \begin{narrow}{-0.3in}{0in}
    \begin{minipage}[t]{.48\linewidth}
      \includegraphics[scale=0.5]{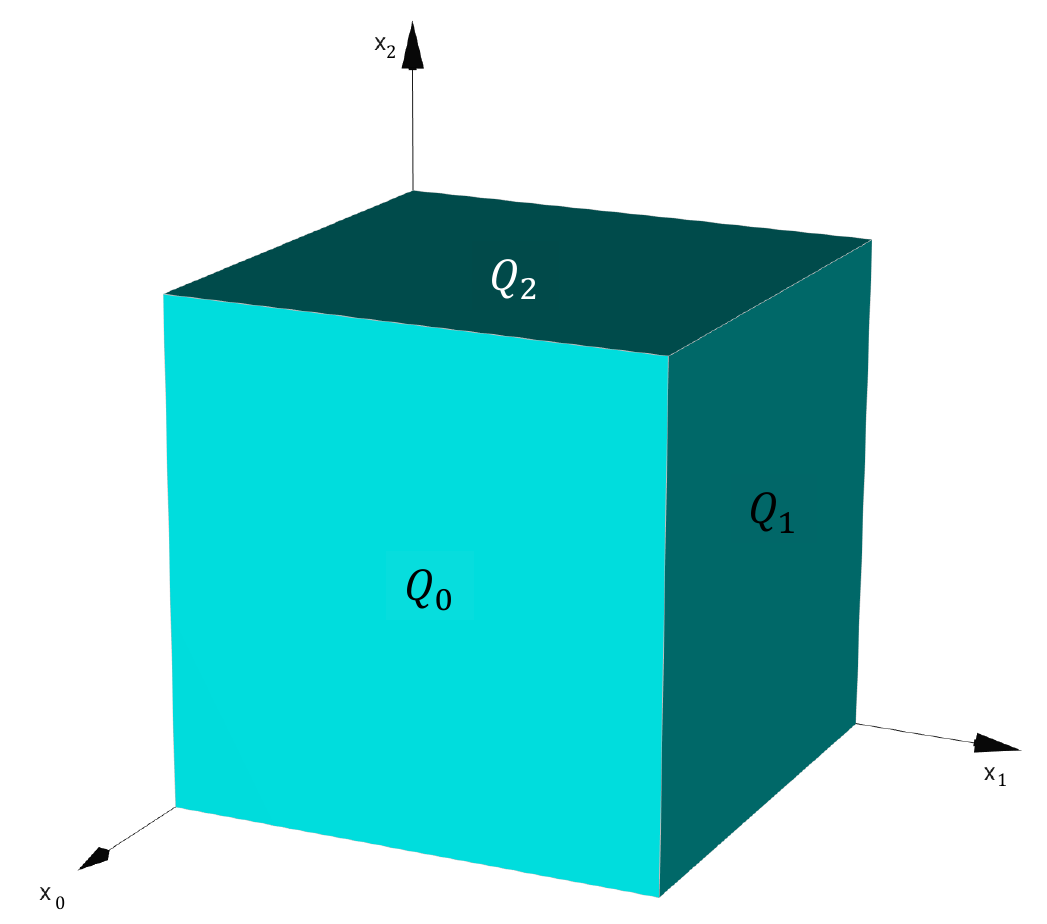}
    \end{minipage}\hspace{2cm}
    \begin{minipage}[t]{.48\linewidth}
      \includegraphics[scale=0.4]{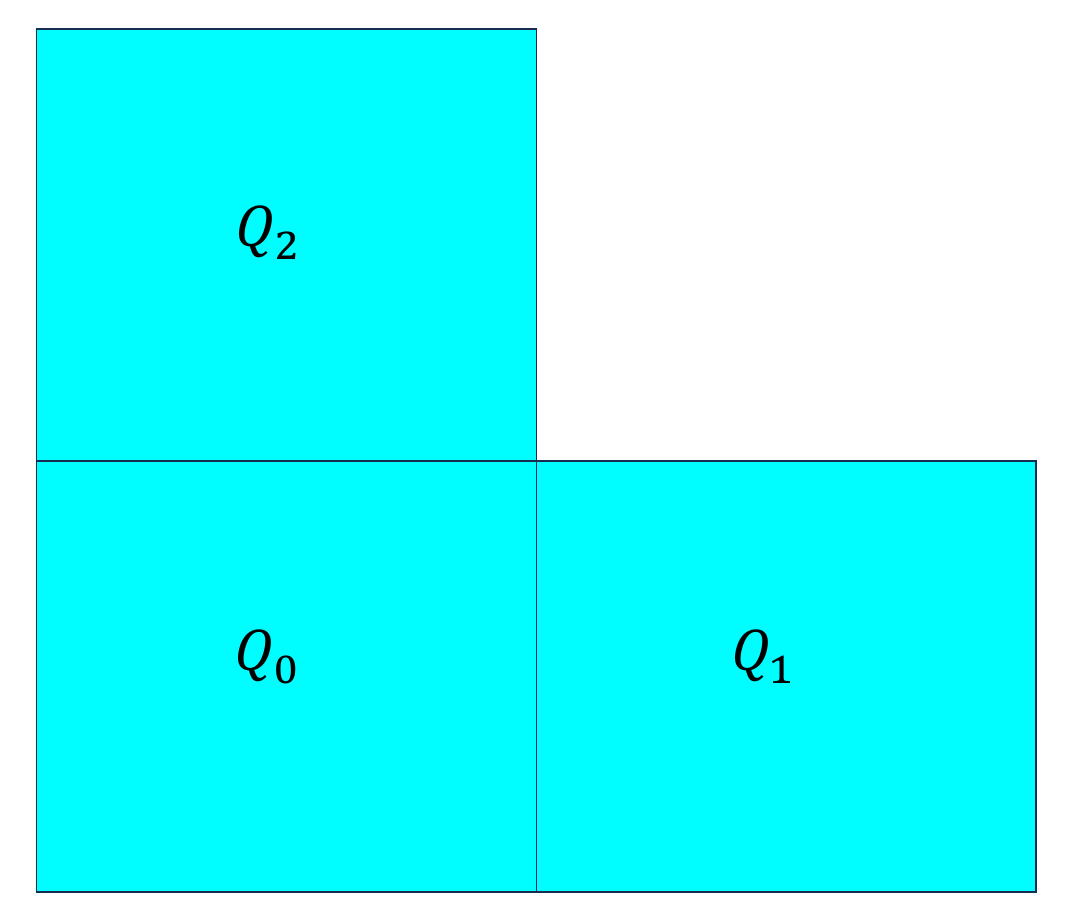}
    \end{minipage}
  \end{narrow}
  \captionof{figure}{Unfolding the cells of \(\Delta^2_{\infty}\) by rotation
    around the edges adjacent to the cell \(Q_0\).\label{reshaping:fig}}
\end{center}

The formula for the rotation of \(Q_{j}\) around the face common with \(Q_{i}\),
so that the rotated \(Q_{j}\) aligns with \(Q_{i}\), assuming \(i < j\), is
\[
  r_{ij}(\ldots, x_{i}, x_{i+1},\ldots, x_{j-1}, 1, \ldots) = (\ldots, 1, x_{i+1}, \ldots, x_{j-1}, 2 - x_{i}, \ldots)
\]
That is, the \(i\)-th coordinate in the rotated \(Q_{j}\) is set to 1 and the
\(j\)-th coordinate is \(2 - x_{i}\).

By stretching the \(L^{\infty}\)-cells adjacent to the \(L^{\infty}\)-cell
\(Q_{i}\) to which the other \(L^{\infty}\)-cells are aligned (\(Q_{i} = Q_{0}\)
in Figure~\ref{reshaping:fig} and reshaped are \(L^{\infty}\)-cells \(Q_{1}\)
and \(Q_{2}\)) we get a global coordinate system
\(r_{i}: \mathbb{RP}^{d}_{\ge 0} \rightarrow [0,2]^{d}\) over
\(\mathbb{RP}^{d}_{\ge 0}\) (see Figure~\ref{reshapeQ2:fig} for the reshaping
process and Figure~\ref{reshape.final:fig} for the final result after stretching
all cells adjacent to the central cell). The map though is not smooth in the
interior of \(\mathbb{RP}^{d}_{\ge 0}\) and in the next section we are going to
show how one can create smooth global coordinate systems over any composition
spaces.

\begin{center}
  \includegraphics[scale=0.5]{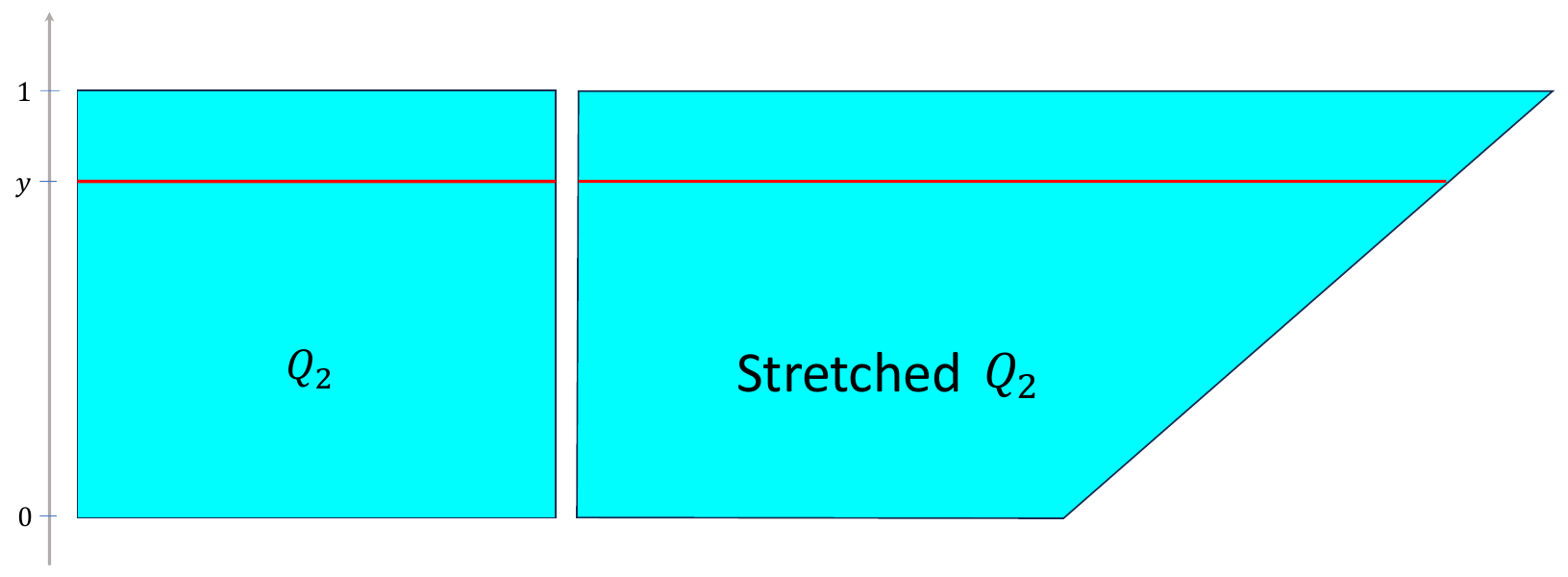}
  \captionof{figure}{Reshaping the \(L^{\infty}\)-cell \(Q_{2}\) of
    \(\Delta^2_{\infty}\) so that after similar reshaping of the
    \(L^{\infty}\)-cell \(Q_{1}\) the resulting regions is \([0,2]^{2}\),
    creating a global (non-smooth) coordinate system over \(\Delta^2_{\infty}\).
    The reshaping the \(L^{\infty}\)-cell \(Q_{2}\) sends a line interval at the
    height \(y\) of \(Q_{2}\) into a line interval in the stretched \(Q_{2}\)
    that is of length \(1+y\). Thus, the stretching transformations has the
    formula \((x,y) \mapsto ((1+y)x, y)\) in the 2d case. \label{reshapeQ2:fig}}
\end{center}

\begin{center}
  \includegraphics[scale=0.5]{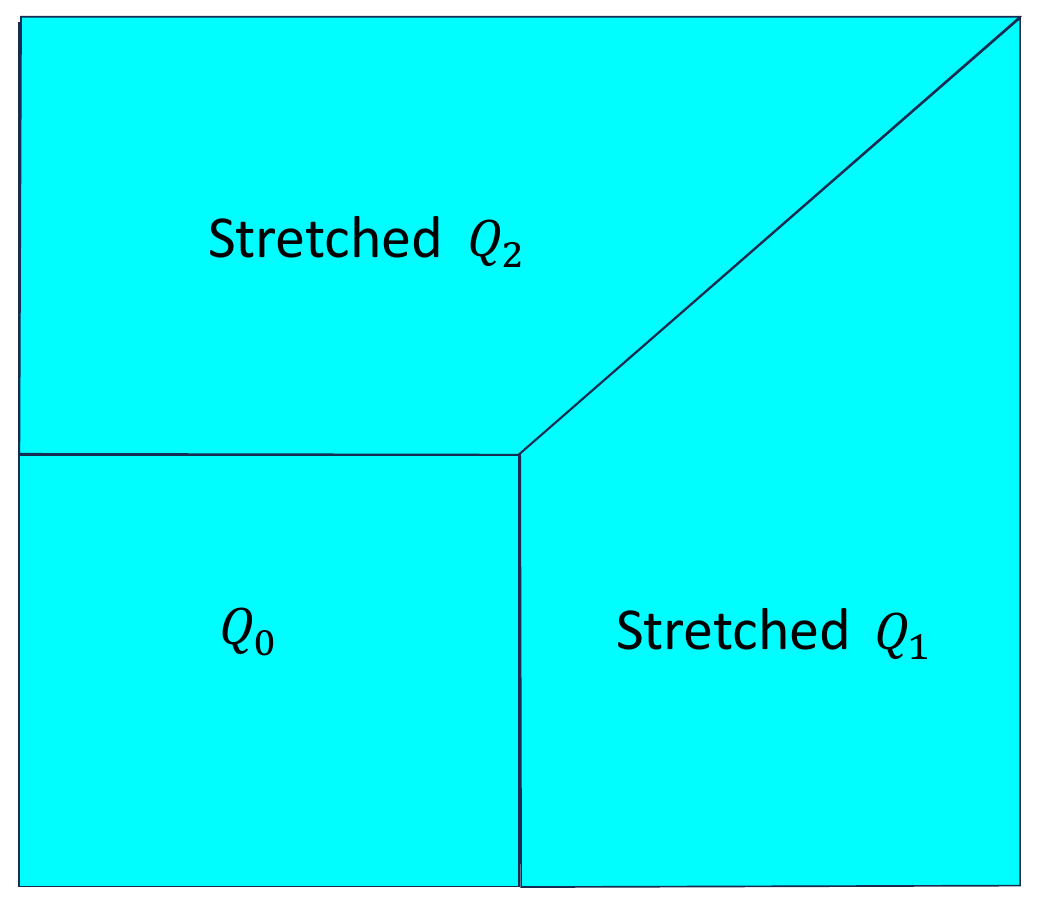}
  \captionof{figure}{Reshaping the cells of \(\Delta^2_{\infty}\) to create a
    global (non-smooth) coordinate system over
    \(\Delta^2_{\infty}\). \label{reshape.final:fig}}
\end{center}

%%% Local Variables:
%%% mode: latex
%%% TeX-master: "Linf_paper"
%%% End:
 % 7
\section*{8.  Hypercube Embeddings of Compositional Data}

\if 0
    This extension is informed by the realization that 1) the image of the
  homogeneous coordinate map can be identified with the set difference,
  \([0,1]^{d} - \Delta^{d-1}_{\infty}\), of the unit cube \([0,1]^{d}\) and the
  \(L^{\infty}\)-simplex \(\Delta^{d-1}_{\infty}\) that occupies a part of the
  boundary of \([0,1]^{d}\) and 2) the set where the homogeneous coordinate map
  is undefined can be identified with \(\Delta^{d-1}_{\infty}\). As a result, we
  arrive at a topologically equivalent form of the homogeneous coordinates map
  \(\mathbb{RP}^{d}_{\ge 0} - \Delta^{d-1}_{\infty} \rightarrow [0,1]^{d} - \Delta^{d-1}_{\infty}\)
  that naturally extends to a homeomorphism
  \(\mathbb{RP}^{d}_{\ge 0} \rightarrow [0,1]^{d}\). The restriction of that
  homeomorphism to a compositional dataset contained in
  \(\mathbb{RP}^{d}_{\ge 0}\) results in a cube embedding of the data into
  \([0,1]^{d}\).

\fi

The \(k\)-th homogeneous coordinate chart, \(\phi_{k}\), induces a homeomorphism
between \(\mathbb{RP}^{d}_{\ge 0} - \{ x_{k} = 0\}\) and \([0, \infty)^{d}\) as
it is a restriction of the the \(k\)-th homogeneous coordinate chart
homeomorphism
\(\mathbb{RP}^{d}_{\ge 0} - \{ x_{k} = 0\} \rightarrow \mathbb{R}^{d}\) to
\(\mathbb{RP}^{d}_{\ge 0}\). Given that the interval \([0, \infty)\) is
homeomorphic to \([0,1)\), the product space \([0, \infty)^{d}\) is likewise
homeomorphic to \([0,1)^{d}\). Consequently, there is a mapping
\[
  \psi \circ \phi_{k}: \mathbb{RP}^{d}_{\ge 0} - \{ x_{k} = 0\} \rightarrow [0, 1)^{d}
\]
where \(\psi\) is a homeomorphism between \([0, \infty)^{d}\) and \([0,1)^{d}\).
A natural question then arises: Is it possible to find \(\psi\) such that
\(\psi \circ \phi_{k}\) extends to a homeomorphism
\(\hat{\phi}^{\psi}_{k}: \mathbb{RP}^{d}_{\ge 0} \rightarrow [0, 1]^{d}\),
effectively providing a global coordinate system for the compositional space.

In this section, we present a simple geometric construct of a homeomorphism
\(r^{\sigma}: [0, \infty)^{d} \rightarrow [0,1)^{d}\), such that the composition
\(r^{\sigma} \circ \phi_{k}\) extends to a hypercube embedding
\(\hat{\phi}^{\sigma}_{k}: \mathbb{RP}^{d}_{\ge 0} \rightarrow [0, 1]^{d}\). The
mapping \(r^{\sigma}\) is a generalization of the sigmoidal map
\(\sigma: [0, \infty) \rightarrow [0,1)\) to higher dimensions, and the
construction of an extension \(\hat{\phi}^{\sigma}_{k}\) is a generalization of
the extension of
\(\sigma \circ \phi_{1}: \mathbb{RP}^{1}_{\ge 0} - \{x_{1} = 0\} \rightarrow [0, 1)\)
to \(\hat{\phi}_{1}^{\sigma}: \mathbb{RP}^{1}_{\ge 0} \rightarrow [0, 1]\)
presented in Example 2-C of Section~2 to higher dimensions.

Before presenting a general construction of a homeomorphism
\(r^{\sigma}: [0, \infty)^{d} \rightarrow [0,1)^{d}\), such that the composition
\(r^{\sigma} \circ \phi_{k}\) extends to a homeomorphism
\(\hat{\phi}^{\sigma}_{k}: \mathbb{RP}^{d}_{\ge 0} \rightarrow [0, 1]^{d}\), we
are going to illustrate the construct in the two-dimensional case. In dimension
two, the value, \(\phi_{1}([x_{0}: x_{1}: x_{2}])\), of the \(2\)-nd homogeneous
coordinate chart \(\phi_{1}\) can be geometrically interpreted as the
intersection \(\big(\frac{x_{0}}{x_{1}}, 1, \frac{x_{2}}{x_{1}}\big)\) of the
line \(\{\lambda (x_{0}, x_{1}, x_{2})\}_{\lambda \in [0,\infty)}\) representing
\([x_{0}: x_{1}: x_{2}]\) with the plane \(x_{1} = 1\) together with the
identification of that plane with \(\R^{2}\) by dropping the \(1\) at the
second coordinate position (see Figure~\ref{phi2.p1:fig}). Since we are
considering the restriction of \(\phi_{1}\) to \(\mathbb{RP}^{d}_{\ge 0}\), the
line \(\{\lambda (x_{0}, x_{1}, x_{2})\}_{\lambda \in [0,\infty)}\) is
intersected with the positive orthant
\(H_{1} = \R^{3}_{\ge 0} \cap \{x_{1} = 1\}\) of the plane \(x_{1} = 1\) that is
homeomorphic with \([0,\infty)^{2}\).
\begin{center}
  \includegraphics[scale=0.5]{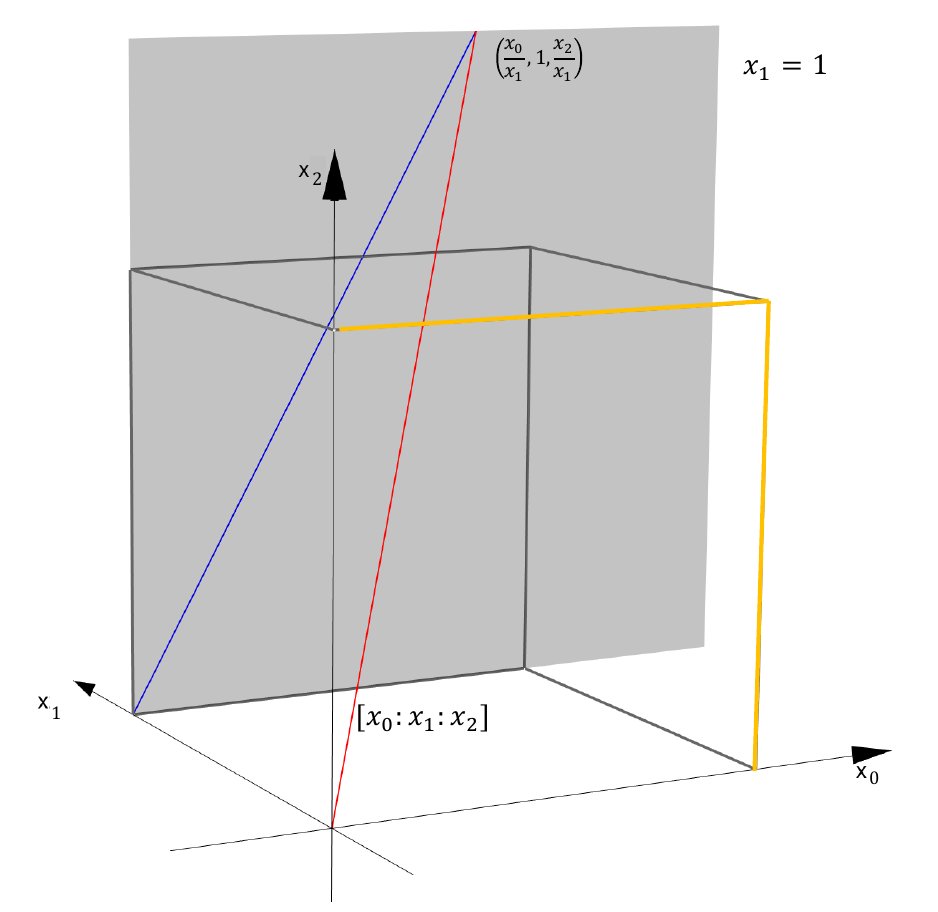}
  \captionof{figure}{Geometric interpretation of \(\phi_{1}\) together with the
    outline of the \(L^{\infty}\)-simplex \(\Delta_{\infty}^{2}\) with the
    subset \(\Delta_{\infty}^{1,2}\) shown in orange of the points of
    \(\Delta_{\infty}^{2}\) contained in the plane \(x_{1} = 0\).
    The shaded area is \(H_{1}\).
    \label{phi2.p1:fig}}
\end{center}

Consider a line that passes through the origin of the positive orthant plane
\(H_{1}\) and the point
\(\big(\frac{x_{0}}{x_{1}}, 1, \frac{x_{2}}{x_{1}}\big)\), shown in
Figures~\ref{phi2.p1:fig}, \ref{phi2.p2:fig}, \ref{phi2.p3:fig} in blue. The
\(L^{\infty}\)-normalization of the line is the projection of that line on
\(\Delta_{\infty}^{2}\) as shown in red in Figure~\ref{phi2.p2:fig}.
\begin{center}
  \includegraphics[scale=0.45]{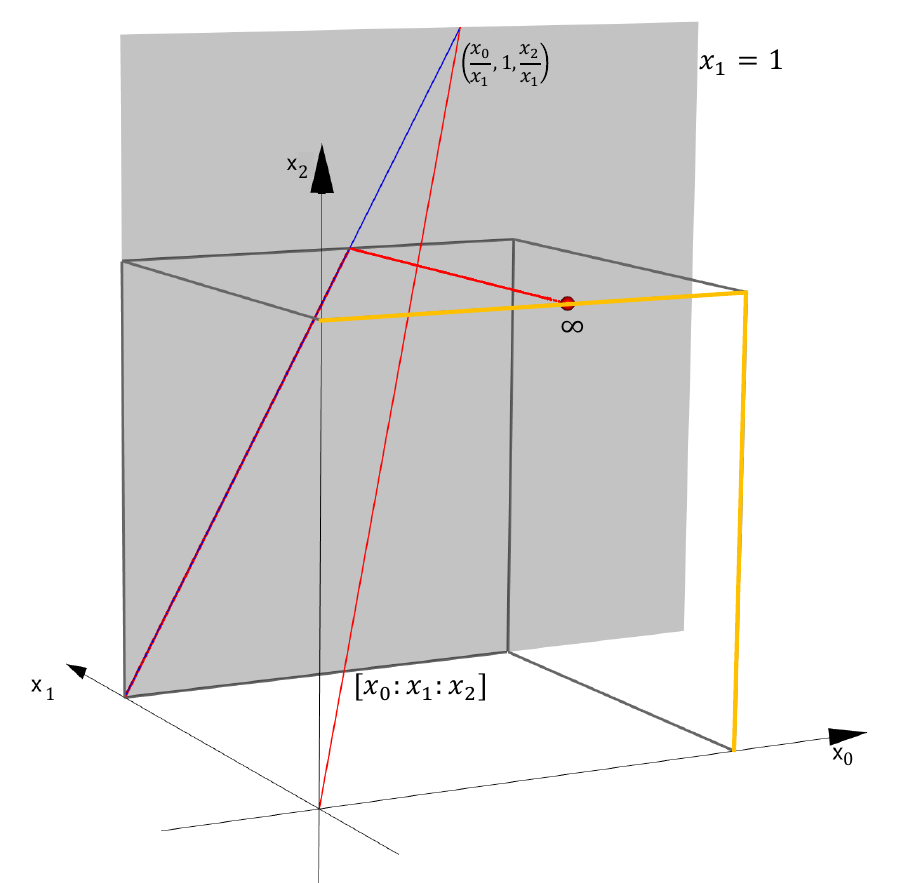}
  \captionof{figure}{Projection of the line (shown in blue) through the origin
    of the plane \(x_{1} = 1\) and the point
    \(\big(\frac{x_{0}}{x_{1}}, 1, \frac{x_{2}}{x_{1}}\big)\) onto
    \(\Delta_{\infty}^{2}\) - shown in red. Together with the point on
    \(\Delta_{\infty}^{1,2} = \{x_{1} = 0\} \cap \Delta_{\infty}^{2}\) shown as
    red sphere that is mapped to a specific point at infinity of the blue line.
     \label{phi2.p2:fig}}
\end{center}
This suggests a mapping of \(H_{1} \cong [0,\infty)^{2}\) onto \([0,1)^{2}\),
interpreted as a subset of the unit square of the plane \(x_{1} = 0\), that
sends any line (shown in blue) through the origin in \(H_{1}\) onto the open
line segment (shown in red in Figure~\ref{phi2.p3:fig}) corresponding to that
line within the unit \(L^{\infty}\)-disk in the plane \(x_{1} = 0\), defined as
the set \(\{(x_{0},0,x_{2}): 0\le x_{i}\le 1, \}\). The mapping is given by the
radial-\(\sigma\) map \(r^{\sigma}\) that maps \(z \in [0,\infty)^{2}\) into
\(\sigma(\| z \|_{1}) \frac{z}{\| z \|_{\infty}} \in [0,1)^{2}\) (see
Figure~\ref{phi2.p3:fig}). In general, the radial-\(\sigma\) map \(r^{\sigma}\)
maps \(z \in [0,\infty)^{d}\) into
\(\sigma(\| z \|_{1}) \frac{z}{\| z \|_{\infty}} \in [0,1)^{d}\).
\begin{center}
  \begin{narrow}{-0.3in}{0in}
    \begin{minipage}[t]{.48\linewidth}
      \includegraphics[scale=0.45]{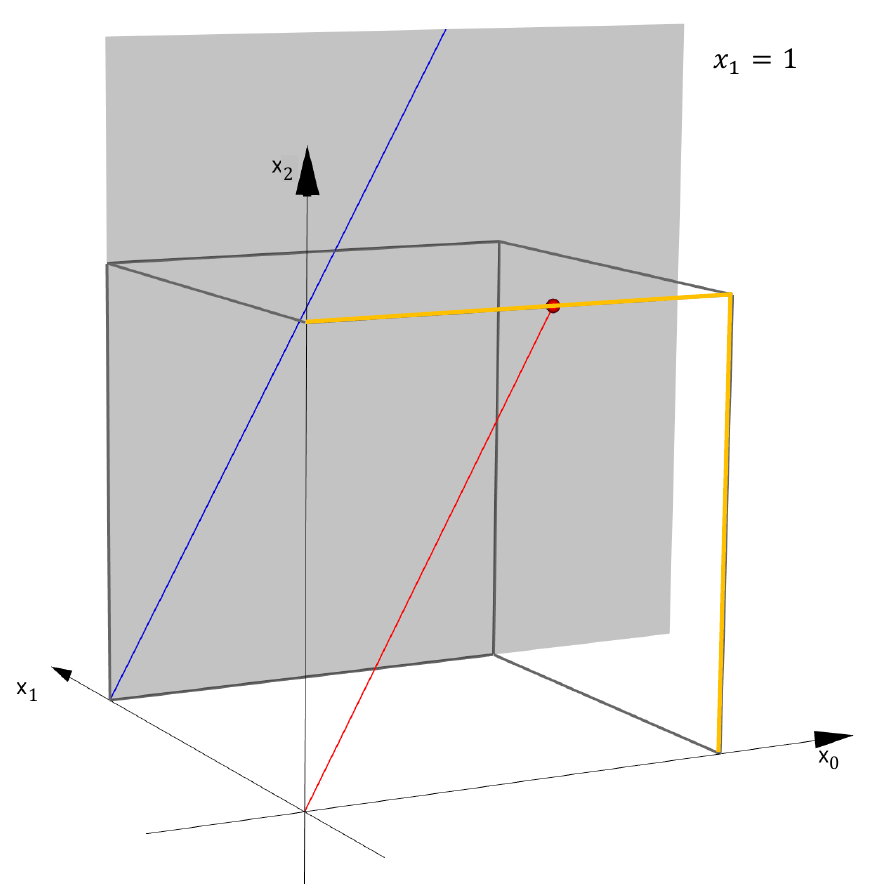}
    \end{minipage}\hspace{-1cm}
    \begin{minipage}[t]{.48\linewidth}
      \includegraphics[scale=0.35]{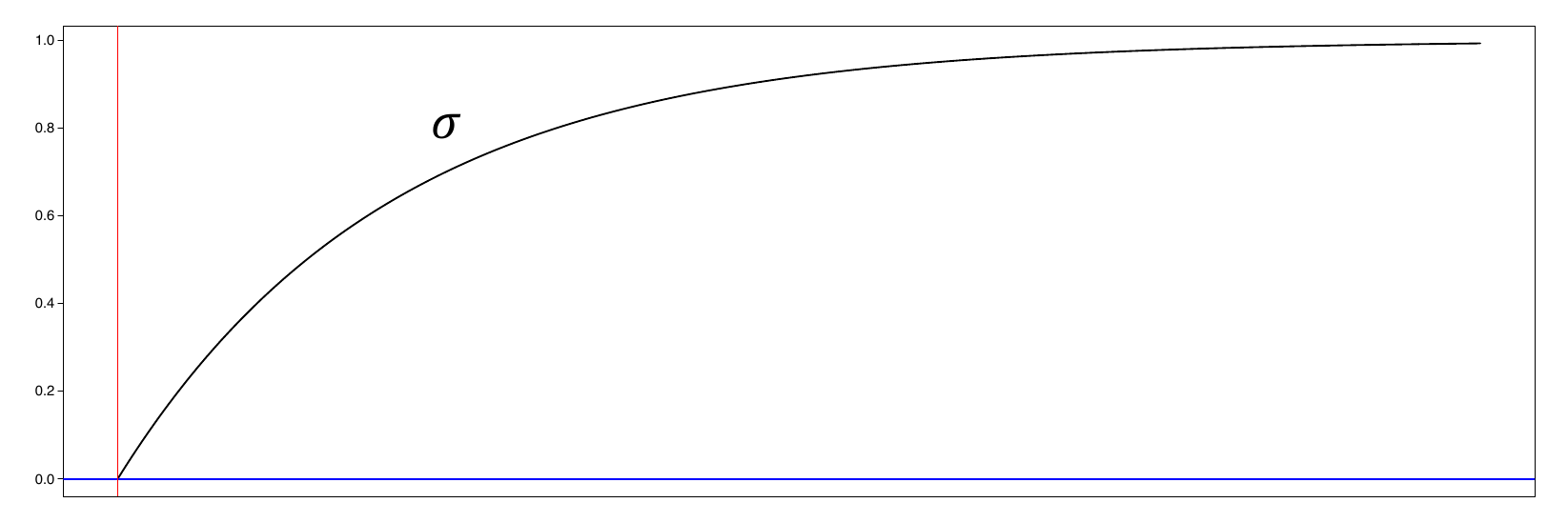}
    \end{minipage}
  \end{narrow}
  \captionof{figure}{Any line through the origin in \(H_{1}\), shown in blue, is
    sent by radial-\(\sigma\) to the fragment of the same line (after parallel
    shift from the \(x_{1} = 1\) plane to the \(x_{1} = 0\) plane) restricted to
    the open unit square \([0,1)^{2}\). \label{phi2.p3:fig}}
\end{center}

In general, the composition \(\phi_{k}^{\sigma} = r^{\sigma} \circ \phi_{k}\) is
defined for \([x]\) in \(\mathbb{RP}^{d}_{\ge 0} - \{ x_{k} = 0\} \) as
\[
  \phi_{k}^{\sigma}([x]) = \sigma(\| \phi_{k}([x]) \|_{1}) \frac{\phi_{k}([x])}{\| \phi_{k}([x]) \|_{\infty}}
\]
The extension \(\hat{\phi}_{k}^{\sigma}\) of \(\phi_{k}^{\sigma}\) to the points
of \(\mathbb{RP}^{d}_{\ge 0}\) contained in \(\{ x_{k} = 0\}\) is defined as
\[
  \hat{\phi}_{k}^{\sigma}([x]) = \frac{x}{\| x \|_{\infty}}
\]
Thus, it is the \(L^{\infty}\)-parametrization of the points of
\(\{ x_{k} = 0\}\). When working with the \(L^{\infty}\)-normalized data,
\(\hat{\phi}_{k}^{\sigma}\) is the identity mapping over \(\{ x_{k} = 0\}\).

\begin{center}
  \includegraphics[scale=0.35]{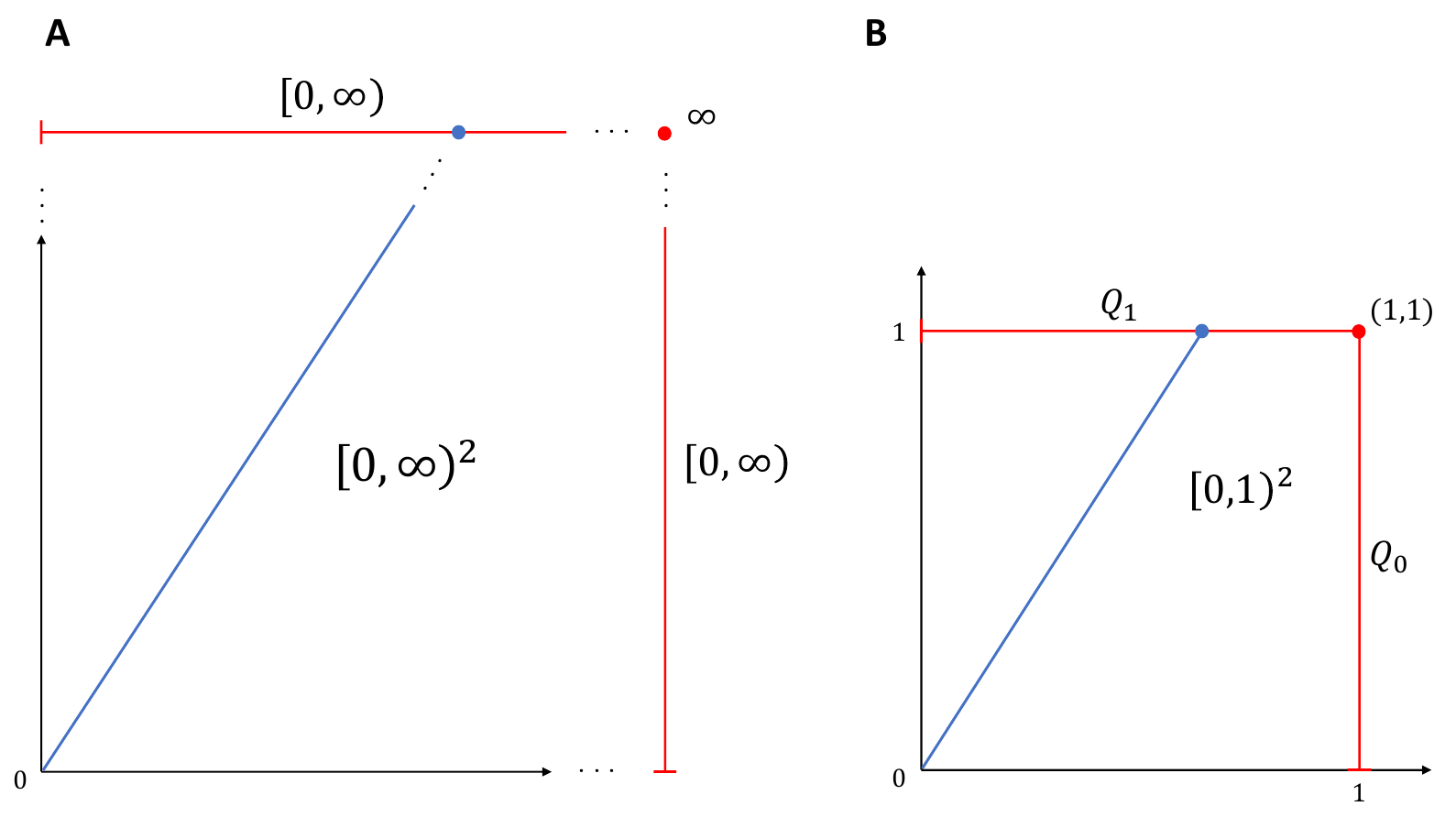}
  \captionof{figure}{Extension of the radial-\(\sigma\) mapping from
    \([0,\infty)^{2} \rightarrow [0,1)^{2}\) to
    \([0,\infty)^{2}* \rightarrow [0,1]^{2}\), where \([0,\infty)^{2}*\) is the
    compactification of \([0,\infty)^{2}\) with the union
    \([0,\infty) \cup \{\infty\} \cup [0,\infty)\), highlighted in red in Panel
    A. Panel B shows the unit square \([0,1]^{2}\) as a compactification of
    \([0,1)^{2}\) with \(\Delta_{\infty}^{1} = Q_{0} \cup Q_{1}\) marked in red.
    The extension \(\hat{r}^{\sigma}\) maps the infinity points of the
    compactification \([0,\infty)^{2}*\) of \([0,\infty)^{2}\) through a
    sigmoidal function onto \(\Delta_{\infty}^{1}\). The radial-\(\sigma\)
    mapping sends the blue line in the panel A into the corresponding blue line
    segment in the panel B, mapping the infinite point on the left (shown as a
    blue disk) into the corresponding blue disk at the end of the corresponding
    line segment. \label{hatr:fig}}
\end{center}

In the two-dimensional case, the hypercube embedding
\(\hat{\phi}^{\sigma}_{k}: \mathbb{RP}^{2}_{\ge 0} \rightarrow [0, 1]^{2}\)
factors out into the composition
\[
  \hat{r}^{\sigma} \circ \phi_{k}^{*}: \mathbb{RP}^{2}_{\ge 0} \rightarrow [0, \infty)^{2}* \rightarrow [0, 1]^{2},
\]
where
\[
  [0, \infty)^{2}* = [0, \infty)^{2} \cup [0, \infty) \cup \{\infty\} \cup [0, \infty)
\]
is a compactification of \([0, \infty)^{2}\) depicted in the panel A of
Figure~\ref{hatr:fig} and \(\hat{r}^{\sigma}\) is an extension of the
radial-\(\sigma\) mapping \(r^{\sigma}\) to a homeomorphism
\([0, \infty)^{2}* \rightarrow [0, 1]^{2}\) mapping the first copy of
\([0, \infty)\) in the compactification of \([0, \infty)^{2}\) to
\([0,1) \subset Q_{0}\) of \(\Delta^{1}_{\infty}\), mapping \(\infty\) to
\((1,1) \in \Delta^{1}_{\infty}\) and mapping the second copy of \([0, \infty)\)
in the compactification of \([0, \infty)^{2}\) to \([0,1) \subset Q_{1}\) of
\(\Delta^{1}_{\infty}\) as illustrated in Figure~\ref{hatr:fig}. For \(k = 2\),
\(\phi_{1}^{*}\) is defined over \(\{x_{1} = 0\}\) as
\[
  \phi_{1}^{*}([x_{0}: 0 : x_{2}]) =
  \begin{cases}
    \sigma^{-1}(\frac{x_{0}}{x_{2}}) & \text{if } x_{0} < x_{2}, \\
    \infty & \text{if } x_{0} = x_{2}, \\
    \sigma^{-1}(\frac{x_{2}}{x_{0}}) & \text{if } x_{0} > x_{2}.
  \end{cases}
\]
In general, one can show that the hypercube embedding
\(\hat{\phi}^{\sigma}_{k}: \mathbb{RP}^{d}_{\ge 0} \rightarrow [0, 1]^{d}\) can
be represented as the composition
\(\hat{r}^{\sigma} \circ \phi_{k}*: \mathbb{RP}^{d}_{\ge 0} \rightarrow [0, \infty)^{d}* \rightarrow [0, 1]^{d}\),
where \([0, \infty)^{d}*\) is a compactification of \([0, \infty)^{d}\) by a set
of points at infinity homeomorphic to \(\Delta_{\infty}^{d-1}\) and
\(\hat{r}^{\sigma}\) is an extension of the radial-\(\sigma\) mapping
\(r^{\sigma}\) to a homeomorphism \([0, \infty)^{d}* \rightarrow [0, 1]^{d}\).
Given that this fact is purely of theoretical nature without practical
implications for analyzing real-world data, we offer the proof of this statement
as a challenge to readers with a penchant for topology.

Implementing the hypercube embedding algorithm necessitates careful management
of rounding errors. To address this, we employ the sigmoidal function
\(\sigma(x,\lambda) = 1 - e^{-\lambda x}\), and for any given dataset, we
determine a value for \(\lambda > 0\) that ensures \(1 - e^{-\lambda M} \ne 1\)
and \(1 - e^{-\lambda m} \ne 0\), where \(M\) represents the maximum and \(m\)
the minimum of \(\| \phi_{k}([x]) \|_{1}\) across the dataset. Specifically, we
aim for \(1 - e^{-\lambda M} < C\varepsilon\) and
\(1 - e^{-\lambda m} > C\varepsilon\), with \(\varepsilon\) being the smallest
double floating-point number for which \(1 + \varepsilon \ne 1\), and \(C > 0\)
fulfilling the condition:
\[
  -\frac{\log(C\varepsilon)}{M} \le  -\frac{\log(1 - C\varepsilon)}{m}.
\]
\begin{center}
\includegraphics[scale=0.11]{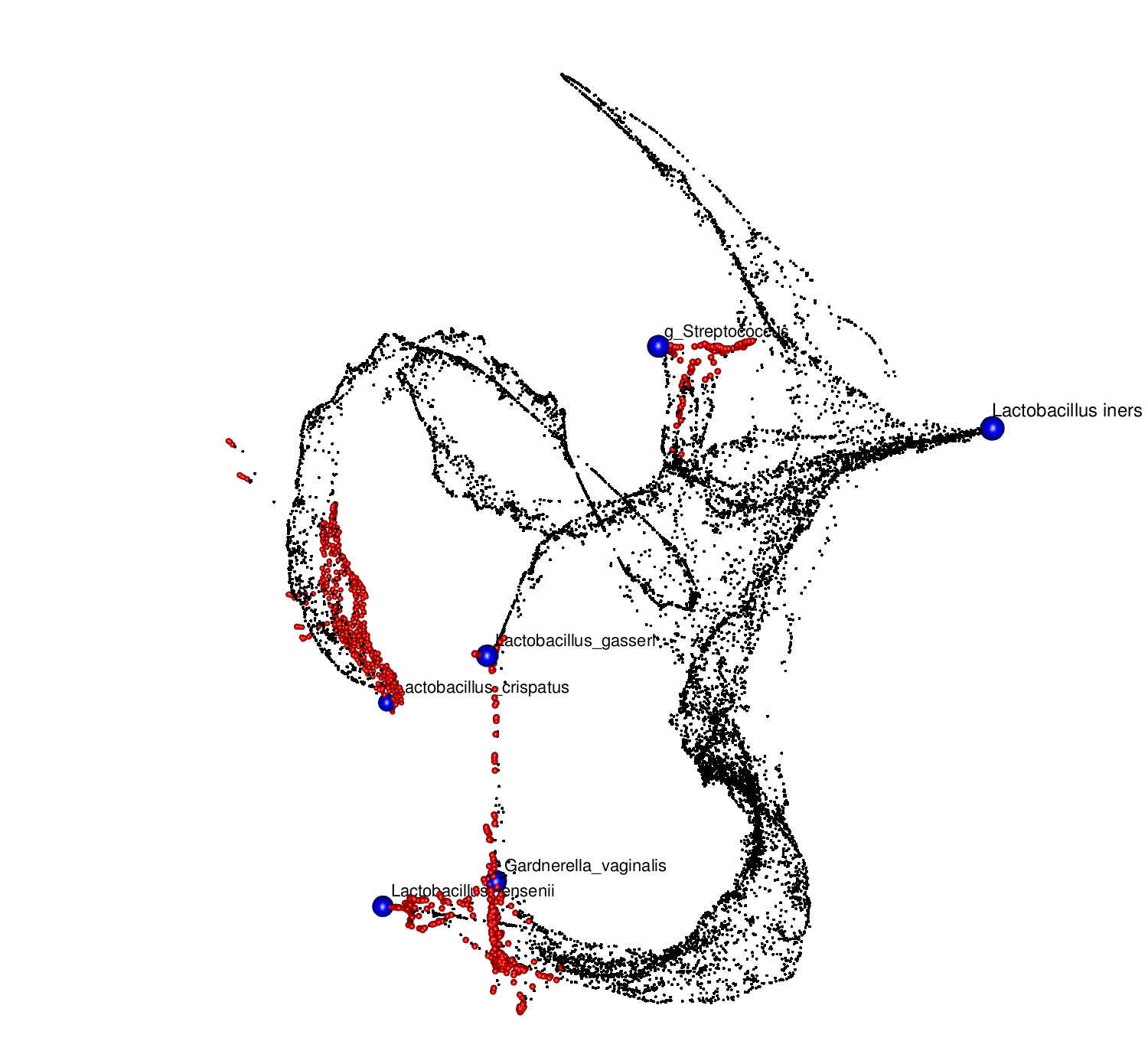}
\captionof{figure}{Low dimensional PaCMAP representation of the Li-hypercube
  embedding of the VALENCIA 16S rRNA
  dataset. The points at infinity are shown in red. \label{S.Li.hce.with.infty.pts.pacmap.3d:fig}}
\end{center}

Figures~\ref{S.Li.hce.with.infty.pts.pacmap.3d:fig},
\ref{S.Lc.hce.with.infty.pts.pacmap.3d:fig} and
\ref{S.Gv.hce.with.infty.pts.pacmap.3d:fig} show PaCMAP 3d representation of the
Li, Lc and Gv, respectively, hypercube embeddings of the VALENCIA 16S rRNA dataset with the
points at infinity shown in red. The points at infinity are smoothly adjacent to
the finite points indicating the continuous nature of the hypercube embedding.

\begin{center}
\includegraphics[scale=0.2]{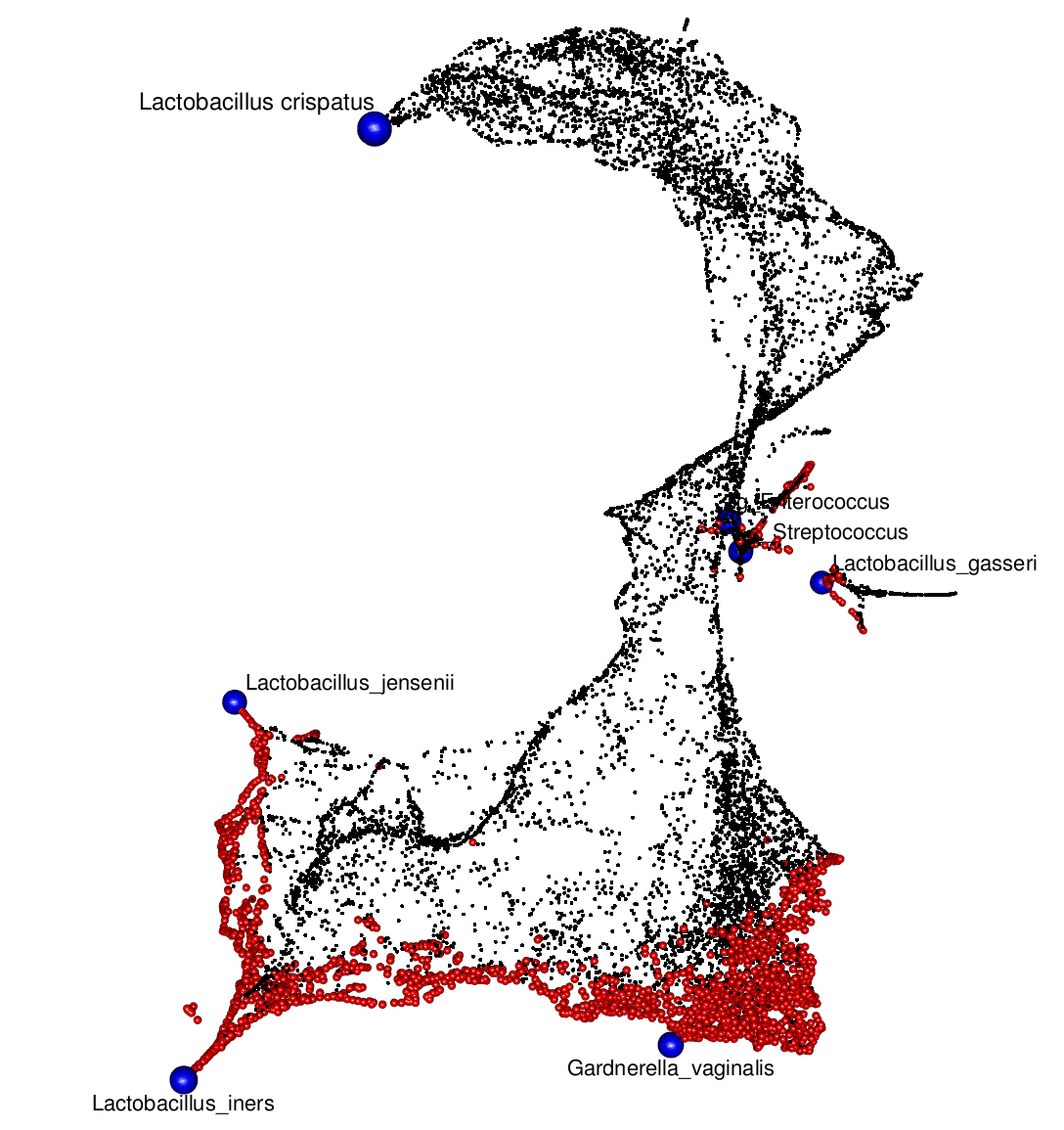}
\captionof{figure}{Low dimensional PaCMAP representation of the Lc-hypercube
  embedding of the VALENCIA 16S rRNA
  dataset. \label{S.Lc.hce.with.infty.pts.pacmap.3d:fig}}
\end{center}

\begin{center}
\includegraphics[scale=0.2]{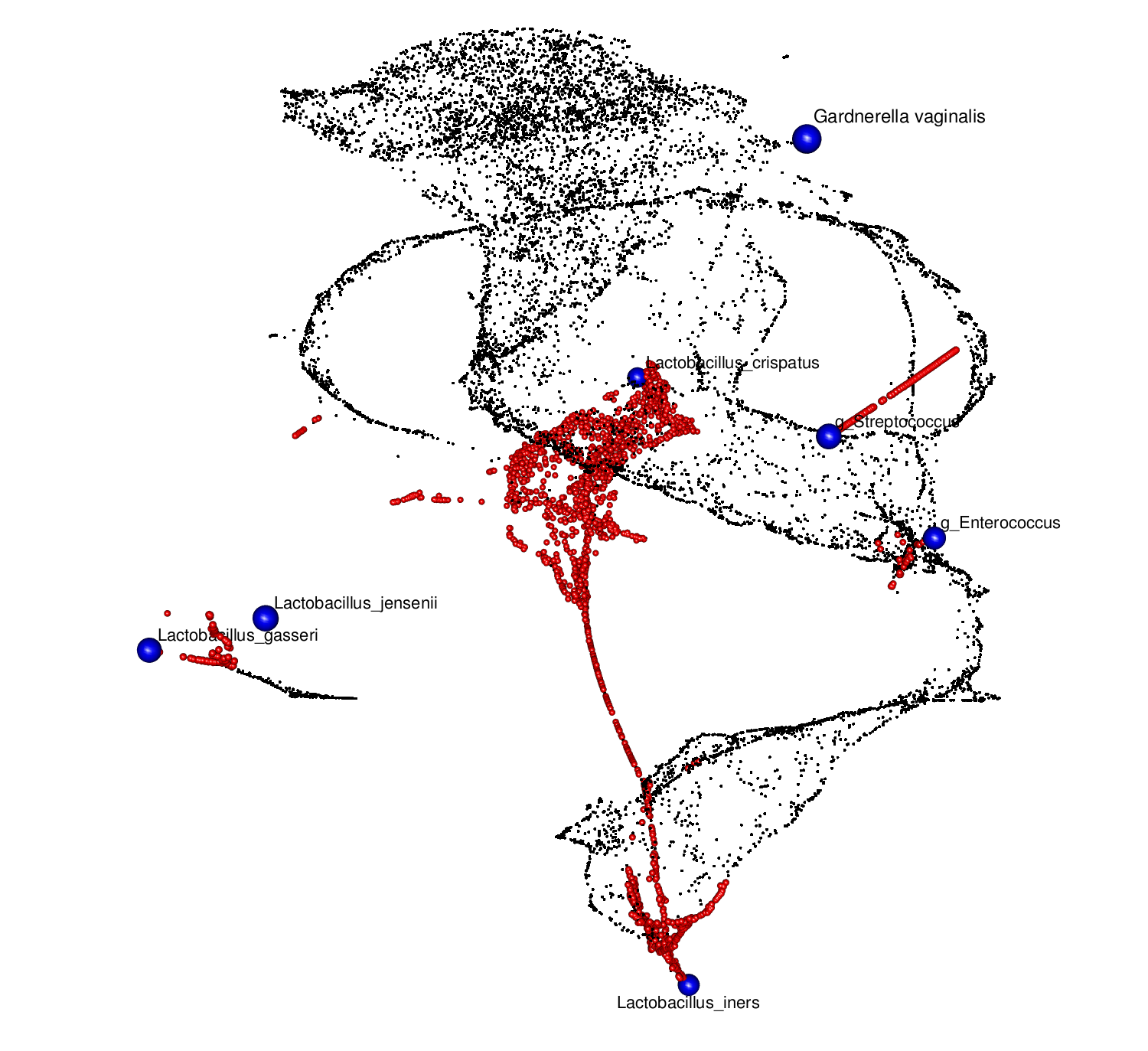}
\captionof{figure}{Low dimensional PaCMAP representation of the Gv-hypercube
  embedding of the VALENCIA 16S rRNA
  dataset. \label{S.Gv.hce.with.infty.pts.pacmap.3d:fig}}
\end{center}

%%% Local Variables:
%%% mode: latex
%%% TeX-master: "Linf_paper"
%%% End:
 % 8
\section*{9. Combining Cube Embeddings}

Cube embeddings allow for study microbial communities from the point of
view of absolute abundance ratios of all components with respect to the given
reference component. Thus, for a given reference component, the cube
embedding of this reference component carries information about the way
abundances of other components relate with the abundance of the reference over
all samples. However, the abundance of representations introduces complexity,
prompting the question: Can these diverse representations be integrated? A
unified representation can be defined as the Cartesian product of the cube
embeddings of the data's \(L^{\infty}\)-CSTs or a subset of \(L^{\infty}\)-CSTs.
The following figure shows the PaCMAP 3d representation of the product embedding
of the combined Li, Lc and Gv-cube embeddings as they correspond to the
largest \(L^{\infty}\)-CSTs.
\begin{center}
  \begin{narrow}{-0.8in}{0in}
    \includegraphics[scale=0.27]{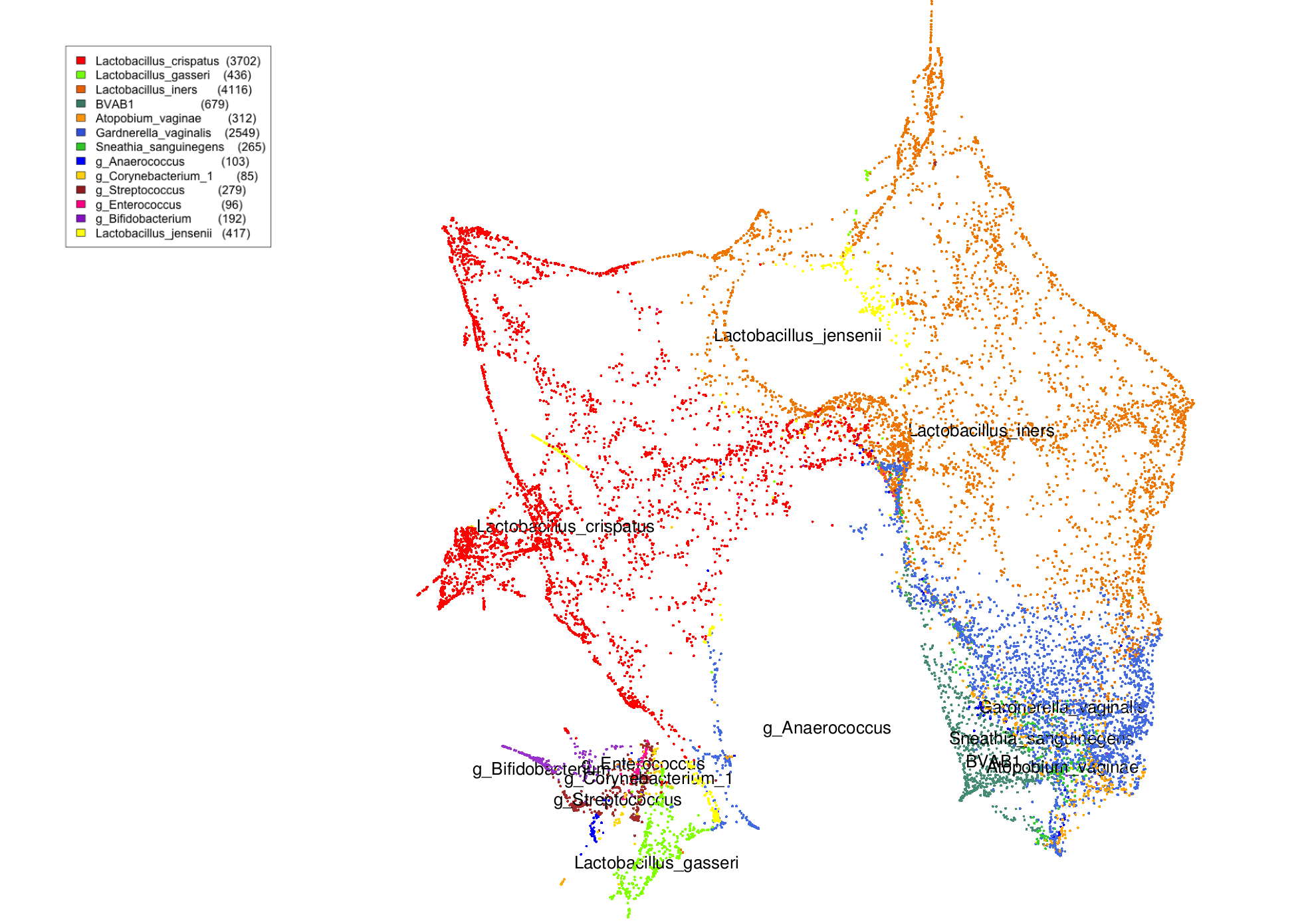}
  \end{narrow}
  \captionof{figure}{PaCMAP 3d embedding of the combined Li, Lc and Gv-cube
  embeddings color coded by \(L^{\infty}\)-CSTs over the VALENCIA 16S rRNA
  data. \label{Scmb3.pacmap3d.with.Linf.CSTs:fig}}
\end{center}

%%% Local Variables:
%%% mode: latex
%%% TeX-master: "Linf_paper"
%%% End:

 % 9
\section*{10. Discussion}

In this paper, we have introduced a novel approach to compositional data
analysis based on \(L^{\infty}\)-normalization and its associated decomposition
of the compositional space into \(L^{\infty}\)-cells. This approach addresses
the challenges posed by the prevalence of zeros in high-throughput omics
datasets, which are inherently compositional. By focusing on
\(L^{\infty}\)-normalization, we have shown that it possesses advantageous
properties, such as subcompositional coherence with respect to the elimination
of low-abundance components, which is particularly relevant for omics data.

The \(L^{\infty}\)-decomposition of the compositional space provides a new
perspective on the characterization of microbial communities. We have introduced
the concept of \(L^{\infty}\)-CSTs, which are derived from the truncated
\(L^{\infty}\)-decomposition and offer several advantages over classical CSTs or
enterotypes. These advantages include a clear and biologically meaningful
definition, stability under the addition or subtraction of samples, and the
provision of a homogeneous coordinate system for each \(L^{\infty}\)-cell,
facilitating the elucidation of its internal structure. The comparison of
\(L^{\infty}\)-CSTs with VALENCIA CSTs in the context of vaginal microbiome data
has demonstrated a high level of concordance, validating the potential of
\(L^{\infty}\)-CSTs as an alternative approach for high-level characterization
of microbial communities.

While CST-like constructs provide a means for conducting CST-based association
analyses, allowing for the comparison of the prevalence of different factors
across distinct regions within the community state space, the segmentation of
this space into CSTs can also be viewed as a limitation. This discretization of
a naturally continuous state space that lacks distinct clusters can be mitigated
by interpreting each \(L^{\infty}\)-cell as an element of a filtration
\(\{Q(r)\}_{r \in [0,1]}\) of a \(d\)-dimensional cube \([0,1]^{d}\), where
\(Q(r)\) is the \(L^{\infty}\)-disk of radius \(r\) centered at the origin of
\([0,1]^{d}\). This approach allows for the analysis of the dependence of the
mean prevalence of specific factors along the boundary \(\partial Q(r)\) of
\(Q(r)\) on \(r\). Furthermore, by identifying clusters of the given data within
each \(\partial Q(r)\) and representing the prevalence of specific factors along
the boundary \(\partial Q(r)\) of \(Q(r)\) as a mixture of the components
corresponding to these clusters, one can study how these mixture components
depend on the distance from the origin of \([0,1]^{d}\), which corresponds to
the community state completely dominated by the reference component. This
analytical setup can be thought of as a special case of a more general Reeb
complex construct, with \(\partial Q(r)\) viewed as the level sets of the
\(L^{\infty}\)-distance from the origin function
\cite{reeb1946points}. The development of these ideas is left for
future research.

Furthermore, we have introduced cube embeddings, which extend homogeneous
coordinates to the entire sample space, allowing for the study of compositional
data from multiple perspectives. By integrating cube embeddings associated with
\(L^{\infty}\)-CSTs, we have shown that a unified representation of the data can
be obtained, providing a comprehensive understanding of the compositional data's
structure.

The methods presented in this paper have broad applicability beyond microbiome
studies and can be applied to any type of compositional data. The geometric and
topological ideas employed in the development of these methods provide a solid
foundation for further advancements in compositional data analysis.

However, there are still several aspects that require further investigation. One
area of interest is the exploration of lower-dimensional \(L^{\infty}\)-cells,
which may reveal patterns of dominance by consortia of bacteria, particularly in
the context of the gut microbiome. Another avenue for future research is the
application of these methods to other omics data types, such as metagenomics and
metabolomics, to assess their performance and potential for uncovering novel
insights.

In conclusion, the \(L^{\infty}\)-normalization approach and its associated
methods presented in this paper offer a promising framework for compositional
data analysis, particularly in the context of high-throughput omics datasets.
The advantages of \(L^{\infty}\)-CSTs, the introduction of cube embeddings, and
the integration of multiple perspectives through the Cartesian product of cube
embeddings provide a comprehensive set of tools for understanding the structure
of compositional data. Further research and application of these methods to
diverse datasets will help to refine and extend their utility in various fields
of study.

%%% Local Variables:
%%% mode: latex
%%% TeX-master: "Linf_paper"
%%% End:
 % 10

\bibliographystyle{plain} % apalike, alpha, plain
\bibliography{pg}{}

\begin{thebibliography}{10}

\bibitem{aitchison1982statistical}
John Aitchison.
\newblock The statistical analysis of compositional data.
\newblock {\em Journal of the Royal Statistical Society: Series B
  (Methodological)}, 44(2):139--160, 1982.

\bibitem{aitchison1986book}
John Aitchison.
\newblock {\em The Statistical Analysis of Compositional Data}.
\newblock Chapman \& Hall, Ltd., GBR, 1986.

\bibitem{france:2020}
Michael France, Bing Ma, Pawel Gajer, Sarah Brown, Mike~S Humphrys, Johanna~B
  Holm, Rebecca~M Brotman, and Jacques Ravel.
\newblock {VALENCIA: A nearest centroid classification method for vaginal
  microbial communities based on composition}.
\newblock {\em Microbiome}, 8(166), 2020.

\bibitem{ge2011data}
Xiaoyin Ge, Issam Safa, Mikhail Belkin, and Yusu Wang.
\newblock Data skeletonization via reeb graphs.
\newblock {\em Advances in neural information processing systems}, 24, 2011.

\bibitem{greenacre2009power}
Michael Greenacre.
\newblock Power transformations in correspondence analysis.
\newblock {\em Computational Statistics \& Data Analysis}, 53(8):3107--3116,
  2009.

\bibitem{guillemin2010differential}
Victor Guillemin and Alan Pollack.
\newblock {\em Differential topology}, volume 370.
\newblock American Mathematical Soc., 2010.

\bibitem{mendelson1990introduction}
Bert Mendelson.
\newblock {\em Introduction to topology}.
\newblock Courier Corporation, 1990.

\bibitem{mobius1827barycentrische}
August~Ferdinand MObius.
\newblock {\em Der barycentrische Calcul ein neues H{\"u}lfsmittel zur
  analytischen Behandlung der Geometrie dargestellt und insbesondere auf die
  Bildung neuer Classen von Aufgaben und die Entwickelung mehrerer
  Eigenschaften der Kegelschnitte angewendet von August Ferdinand Mobius
  Professor der Astronomie zu Leipzig}.
\newblock Verlag von Johann Ambrosius Barth, 1827.

\bibitem{o2020asymptomatic}
D~Elizabeth O'Hanlon, Pawel Gajer, Rebecca~M Brotman, and Jacques Ravel.
\newblock Asymptomatic bacterial vaginosis is associated with depletion of
  mature superficial cells shed from the vaginal epithelium.
\newblock {\em Frontiers in cellular and infection microbiology}, 10:106, 2020.

\bibitem{o2006elementary}
Barrett O'neill.
\newblock {\em Elementary differential geometry}.
\newblock Elsevier, 2006.

\bibitem{ravel2011vaginal}
Jacques Ravel, Pawel Gajer, Zaid Abdo, G~Maria Schneider, Sara~SK Koenig,
  Stacey~L McCulle, Shara Karlebach, Reshma Gorle, Jennifer Russell, Carol~O
  Tacket, et~al.
\newblock Vaginal microbiome of reproductive-age women.
\newblock {\em Proceedings of the National Academy of Sciences},
  108(supplement\_1):4680--4687, 2011.

\bibitem{reeb1946points}
Georges Reeb.
\newblock On the singular points of a completely integrable pfaff form or of a
  numerical function].
\newblock {\em Comptes Rendus Acad. Sciences Paris}, 222:847--849, 1946.

\bibitem{romero2014vaginal}
Roberto Romero, Sonia~S Hassan, Pawel Gajer, Adi~L Tarca, Douglas~W Fadrosh,
  Janine Bieda, Piya Chaemsaithong, Jezid Miranda, Tinnakorn Chaiworapongsa,
  and Jacques Ravel.
\newblock The vaginal microbiota of pregnant women who subsequently have
  spontaneous preterm labor and delivery and those with a normal delivery at
  term.
\newblock {\em Microbiome}, 2:1--15, 2014.

\bibitem{tuddenham2019associations}
Susan Tuddenham, Khalil~G Ghanem, Laura~E Caulfield, Alisha~J Rovner, Courtney
  Robinson, Rupak Shivakoti, Ryan Miller, Anne Burke, Catherine Murphy, Jacques
  Ravel, et~al.
\newblock Associations between dietary micronutrient intake and
  molecular-bacterial vaginosis.
\newblock {\em Reproductive health}, 16(1):1--8, 2019.

\end{thebibliography}

\end{document}